\documentclass{article}

\usepackage{graphicx}
\usepackage{amsthm,amsmath,amssymb} 
\usepackage{color}
\usepackage[mathscr]{euscript}
\usepackage[noadjust]{cite}
\usepackage{multirow}
\usepackage{tabu,array,soul}
\usepackage[a4paper, total={5.5in, 7.5in}]{geometry}
\usepackage{subcaption}
\usepackage{caption}
\usepackage{tikz}
\usepackage{tikz}
\usepackage{rotating}
\usepackage{pdflscape}
\usepackage{lineno}
\usepackage{soul}

\newtheorem{theorem}{Theorem}
\newtheorem{observation}{Observation}

\renewcommand*{\theobservation}{\Alph{observation}}

\newtheorem{corollary}[theorem]{Corollary}
\newenvironment{remark}{\refstepcounter{observation}\medskip\noindent\textbf{Remark \theobservation. }}{\medskip}
\newenvironment{claim}{\medskip\noindent\textit{Claim. }\itshape}{\medskip}
\newtheorem{lemma}[theorem]{Lemma}
\newtheorem{question}{Question}

\newcounter{example3}
\newtheorem{open-question}[example1]{Open Question}
\newtheorem{subclaim}[example3]{Subclaim}

\newcommand{\com}[1]{}

\begin{document}
\bibliographystyle{plain}
\title{On bounds on bend number of split and cocomparability graphs}
\date{}
\author{Dibyayan Chakraborty\thanks{Indian Statistical Institute, Kolkata, India. E-mail: \texttt{dibyayancg@gmail.com}}\and Sandip Das\thanks{Indian Statistical Institute, Kolkata, India. E-mail: \texttt{sandipdas@isical.ac.in}}\and Joydeep Mukherjee\thanks{Indian Statistical Institute, Kolkata, India. E-mail: \texttt{joydeep.m1981@gmail.com}}\and Uma kant Sahoo\thanks{Indian Statistical Institute, Kolkata, India. E-mail: \texttt{umakant.iitkgp@gmail.com}}}

\maketitle

	\begin{abstract}
		A \emph{path} is a simple, piecewise linear curve made up of alternating horizontal and vertical line segments in the plane. A \emph{$k$-bend path} is a path made up of at most $k + 1$ line segments. A \emph{$B_k$-VPG representation} of a graph is a collection of $k$-bend paths such that each path in the collection represents a vertex of the graph and two such paths intersect if and only if the vertices they represent are adjacent in the graph. The graphs that have a $B_k$-VPG representation are called \emph{$B_k$-VPG graphs}. In this paper, we study the relationship between poset dimension and bend number of cocomparability graphs. It is known that the poset dimension $dim(G)$ of a cocomparability graph $G$ is greater than or equal to its bend number $bend(G)$. Cohen et al. ({\textsc{order 2015}}) asked for examples of cocomparability graphs with low bend number and high poset dimension. We answer this question by proving that for each $m, t \in \mathbb{N}$, there exists a  cocomparability graph $G_{t,m}$ with $t < bend(G_{t,m}) \leq 4t+29$ and $dim(G_{t,m})-bend(G_{t,m})>m$. Techniques used to prove the above result, allows us to partially address the open question posed by Chaplick et al. ({\textsc{wg 2012}})  who asked whether $B_k$-VPG-chordal $\subsetneq$ $B_{k+1}$-VPG-chordal for all $k \in \mathbb{N}$. We address this by proving that there are infinitely many $m \in \mathbb{N}$ such that $B_m$-VPG-split $\subsetneq$ $B_{m+1}$-VPG-split which provides infinitely many positive examples. We use the same techniques to prove that, for all $t \in \mathbb{N}$, $B_t$-VPG-$Forb(C_{\geq 5})$ $\subsetneq$ $B_{4t+29}$-VPG-$Forb(C_{\geq 5})$, where $Forb(C_{\geq 5})$ denotes the family of graphs that does not contain induced cycles of length greater than 4. Furthermore, we show that for all $t \in \mathbb{N}$, $PB_t$-VPG-split $\subsetneq PB_{36t+80}$-VPG-split, where $PB_t$-VPG denotes the class of  graphs with proper bend number at most $t$.

	\end{abstract}

	\section{Introduction}
	
	We study the \emph{bend number} of some subclasses of \emph{string} graphs. A graph $G$ has vertex set $V(G)$ and edge set $E(G)$. A \emph{string representation} of a graph is a collection of simple curves on the plane such that each curve in the collection represents a vertex of the graph and two curves intersect if and only if the vertices they represent are adjacent in the graph. The graphs that have a string representation are called \emph{string graphs}. The concepts of \emph{bend number} and \emph{$B_k$-VPG graphs} (introduced by Asinowski et al.~\cite{asinowski2012}) become useful in gaining a better understanding of subclasses of string graphs. A \emph{path} is a simple, piecewise linear curve made up of alternating horizontal and vertical line segments in the plane. A \emph{$k$-bend path} is a path made up of at most  $k + 1$ line segments. A \emph{$B_k$-VPG representation} of a graph is a collection of $k$-bend paths such that each path in the collection represents a vertex of the graph, and two such paths intersect if and only if the vertices they represent are adjacent in the graph. The graphs that have a $B_k$-VPG representation are called \emph{$B_k$-VPG graphs} and the set of all $B_k$-VPG graphs are denoted simply by $B_k$-VPG. A graph is said to be a \emph{VPG graph} if it is a $B_k$-VPG graph for some $k$. The \emph{bend number} of a graph $G$, denoted by $bend(G)$, is the minimum integer $k$ for which $G$ has a $B_k$-VPG representation. Asinowski et al.~\cite{asinowski2012} showed that the family of VPG graphs are equivalent to the family of string graphs. Therefore,  bend number of a  string graph is finite.

	For past several decades, string graphs have been a popular research topic~\cite{krato1991,krato1991exp,schaefer2003,fox2012}. Kratochv\'{\i}l~\cite{krato1991} proved that it is NP-hard to determine whether a graph is a string graph and there are infinite number of graphs which are not string graphs. Since then, researchers have shown that many important families of graphs are string graphs (for example planar graphs, cocomparability graphs, chordal graphs, co-chordal graphs, circular arc graphs~\cite{goncalves2017,krato1991inf}). However, the nature of string representations of such graphs may vary from each other. For example, Matou\v{s}ek and Kratochv\'{\i}l~\cite{krato1991exp} proved the existence of graphs whose any string representation contains an exponential number of intersection points, whereas all planar graphs and circle graphs have a string representation where any two curve intersects at most once~\cite{asinowski2012,biedl2015}. There are some algorithmic results for various classes of $B_k$-VPG graphs. Since graph classes like circle graphs and planar graphs have bounded bend  number~\cite{asinowski2012,biedl2015,goncalves2017}, problems like \textsc{Minimum     Clique Cover}, \textsc{Colorability}, \textsc{Minimum Dominating Set}, \textsc{Hamiltonian Cycle} and \textsc{Maximum independent set} remain \textsc{NP}-hard on $B_k$-VPG graphs for $k\geq 1$. Lahiri et al.~\cite{lahiri2015} gave an approximation algorithm for \textsc{Maximum independent set} in $B_1$-VPG graphs and Mehrabi~\cite{mehrabi2017} gave an approximation algorithm for \textsc{Minimum Dominating Set} in $B_1$-VPG graphs.

	The first result in this paper is about \emph{poset dimension} and bend number of \emph{cocomparability graphs}.
	Cocomparability graphs are an important subclass of string graphs. Consider a poset $P=(X,\preceq)$ where $X$, the \emph{ground set}, is a set of elements and $\preceq$ denotes the \emph{partial order} relation. For two distinct elements $x,y\in X$, we write $x\prec y$ if $x\preceq y$ in $P$. Since the posets considered in this article have distinct elements we can safely replace ``$\preceq$'' by ``$\prec$''. A \emph{linear extension} $L$ on $X$ is a linear order on $X$ such that if $x\prec y$ in $P$ then $x\prec y$ in $L$. A set of linear orders $\cal{L}$ is said to be a \emph{realizer} of $P$ if $P=\cap \cal{L}$ that is $x\prec y$ in $P$ if and only if $x\prec y$ in all linear orders in $\cal{L}$. The \emph{size} of a realizer is the number of linear orders in that realizer. The \emph{dimension} of a poset $P$, denoted by $dim(P)$, is the minimum integer $t$ such that there is a realizer of $P$ with size $t$.
	
	For a poset $P=(X,\preceq)$, the \emph{comparability} graph of $P$ is the undirected graph $G_P$ such that $V(G_P)=X$ and for two vertices $u,v\in V(G_P)$ we have $uv\in E(G_P)$ if and only if $u \prec v$ or $v\prec u$ in $P$. A graph $G$ is a \emph{cocomparability} graph if the complement of $G$, denoted by $\overline{G}$, is a comparability graph. It is known that the poset dimension is comparability invariant~\cite{cohen2016}, meaning that all transitive orientations~\cite{cohen2016} of a given comparability graph have the same dimension. Thus, we may extend the definition of the poset dimension to the cocomparability graph as follows. The \emph{poset dimension}, $dim(G)$, of a cocomparability graph $G$ is the dimension of any poset $P$ for which $\overline{G}$ is isomorphic to $G_P$.
	
	Cohen et al.~\cite{cohen2016} proved that the bend number of a cocomparability graph is strictly less than its poset dimension and initiated the study of the gap between the bend number and poset dimension. They proved the existence of cocomparability graphs with high bend number and high poset dimension. They also provided a class of graphs whose bend number is one but poset dimension is high. They asked the following.
	
	\medskip
	
	\noindent\fbox{
		\parbox{0.96\textwidth}{
			\begin{open-question}    
				Find additional examples of cocomparability graphs with low bend number and high poset dimension.
			\end{open-question}
		}
	}
	\medskip

	We answer this question in affirmative by proving the following stronger theorem.
	
	\begin{theorem}\label{thm:cocomparability}
		\sloppy For each $t, m \in \mathbb{N}$, there exists a cocomparability graph $G_{t,m}$ with $t < bend(G_{t,m}) \leq 4t+29$ and $dim(G_{t,m})-bend(G_{t,m})$ is greater than $m$.
	\end{theorem}

	Asinowski et al.~\cite{asinowski2012} asked whether for every $k\in \mathbb{N}$, $B_{k}$-VPG $\subsetneq B_{k+1}$-VPG. Chaplick et al.~\cite{chaplick2012} answered the above question by constructing a graph which is a $B_{k+1}$-VPG graph but not a $B_k$-VPG graph and proved that it is \textsc{NP}-hard to separate the two classes. Since their constructed graphs contained large cycles, they asked the following open question.
	
	\medskip
	
	\noindent\fbox{
		\parbox{0.96\textwidth}{
			\begin{open-question}\label{open-quest-1}
				Is it true that,  $\forall k\in \mathbb{N}$, $B_{k}$-VPG-chordal $\subsetneq B_{k+1}$-VPG-chordal ?
			\end{open-question}
		}
	}
	\medskip
	
	A graph is chordal if it contains no induced cycles on at least four vertices. Split graphs are those graphs whose vertex set can be partitioned into a clique and independent set. Notice that, any split graph is also a chordal graph.

	Although Open Question~\ref{open-quest-1} is combinatorial in nature, we believe that studying it will give us a better insight into the structure of split and chordal graphs. This, in turn, will lead to the development of better algorithms for these classes of graphs. For example, there have been works on recognising $B_0$-VPG chordal graphs. Chaplick et al.~\cite{chaplick2011recognizing} gave a polynomial time algorithm to recognise $B_0$-VPG chordal graphs. Again it is not difficult to verify that given a $B_0$-VPG representation of a split graph $G$, a minimum dominating set of $G$ can be found in linear time. (To see this consider a split graph $G$ with independent set partition $I$. Observe that, if $G$ is a $B_0$-VPG split graph then its vertex set can be partitioned into two sets $V_1$ and $V_2$ such that both the subgraphs of $G$ induced by $V_1$ and $V_2$ are interval graphs and for any vertex $v\in I$, all the vertices adjacent to $v$ are either subset of $V_1$ or subset of $V_2$. Then we get a minimum dominating set by taking disjoint union of minimum dominating sets of the interval graphs induced on $V_1$ and $V_2$.) So it seems that many difficult problems for split graphs might have efficient algorithms if the bend number of the graph is low.

	With the above motivations, we study the bend number of split graphs. The following theorem partially addresses Open Question~\ref{open-quest-1}.
	
	\begin{theorem}\label{thm:split}
		There exists infinitely many $m \in \mathbb{N}$ such that $B_m$-VPG-split $\subsetneq$ $B_{m+1}$-VPG-split.
	\end{theorem}
	
	
	It is known that there are triangle-free graphs that are not string graphs~\cite{matousek2014}. However, graphs with no induced cycle of length greater than three (i.e. chordal graphs) are string graphs. Therefore it seems that the length of the largest induced cycle in a graph is important in the context of string graphs. Hence we study the bend number of string graphs with no induced cycle of length greater than four. Let $Forb(C_{\geq 5})$ be the class of graphs with no induced cycle of length greater than four. We prove the following.
	
	\begin{theorem}\label{thm:holefree}
		For all $t \in \mathbb{N}$, $B_t$-VPG-$Forb(C_{\geq 5})$ $\subsetneq$ $B_{4t+29}$-VPG-$Forb(C_{\geq 5})$.
	\end{theorem}

	For a given $k\in \mathbb{N}$, there can be graphs whose any $B_k$-VPG representation contains an infinite number of intersection points (when two segments overlap). For example, consider a $B_0$-VPG representation of a complete graph on five vertices. Inspired by Chaplick et al.~\cite{chaplick2012}, we present the concept of \emph{proper bend number} as follows. A $B_k$-VPG representation $\mathcal{R}$ of a graph is a \emph{proper $B_k$-VPG} representation or simply a \emph{$PB_k$-VPG} representation if two paths in $\mathcal{R}$ have only finitely many intersection points, each intersection point belongs to exactly two paths, and whenever two paths intersect they cross each other. The graphs that have a $PB_k$-VPG representation are called \emph{$PB_k$-VPG graphs}. The \emph{proper bend number} of a graph $G$, denoted by $bend_p(G)$, is the minimum integer $k$ for which $G$ has a $PB_k$-VPG representation. From the results of Asinowski et al.~\cite{asinowski2012}, it follows that circle graphs are $PB_1$-VPG graphs and from the results of Biedl et al.~\cite{biedl2015}, it follows that planar graphs have $PB_2$-VPG representation where any two paths intersect at most once. In this paper, we study the proper bend number of split graphs and prove the following theorem.
	
	\begin{theorem}\label{thm:split-2}
		For all $t \in \mathbb{N}$, $PB_t$-VPG-split $\subsetneq PB_{36t+80}$-VPG-split.
	\end{theorem}

	Here, to prove the lower bound of proper bend number of specific split graphs, we give a different technique than used in the proof of Theorem~\ref{thm:split}.
	
	We prove Theorems~\ref{thm:cocomparability},~\ref{thm:split},~\ref{thm:holefree} and \ref{thm:split-2} in Sections~\ref{sec:proof-1},~\ref{sec:proof-2},~\ref{sec:proof-3} and \ref{sec:proof-4} respectively. For better readability, we delay the proofs of some lemmas, concerning upper bounds till Section~\ref{sec:upber-bound}. Finally we conclude in Section~\ref{sec:conclude}.

	\section{Proof of Theorem~\ref{thm:cocomparability}}\label{sec:proof-1}
	
	Proof of Theorem~\ref{thm:cocomparability} is divided into three parts. In Section~\ref{subsec-posdim}, we prove a lower bound for poset dimension of a particular class of posets denoted as $P(r,s;n)$. Then in Section~\ref{subsec-lower-bend} we define a specific class of graphs denoted as $\mathcal{H}_{n,k}$ and prove both lower and upper bounds for its bend number. Finally, in Section~\ref{subsec-combine}, we combine the above results to complete the proof of Theorem~\ref{thm:cocomparability}.

	\subsection{Lower bound on Poset Dimension}\label{subsec-posdim}
	For integers $r$, $s$ and $n$ with $1\leq r < s \leq n-1$, let $T_r$ denote the set of all $r$-element subsets of $[n]$ and $U_{s}$ denote the set of all $s$-element subsets of $[n]$. Let $Y=T_r\cup U_{s}$ and for any two elements $\mu,\tau\in Y$ we say $\mu$ is related to $\tau$, denoted $\mu\supset \tau$, if $\mu$ is a superset of $\tau$. Observe that the relation  $\supset$ along with the set $Y$ induces a poset which is denoted by $P(r,s;n)$.
	
	Consider the poset $P(r,s-1;s)$, for $1 \leq r < s-1$, with realizer $\mathcal{L}$. For a linear order $L\in \mathcal{L}$, we denote the \emph{pivot} element of $L$ as the element $\mu$ such that for any $\mu'\in U_{s-1}\setminus\{\mu\}$, we have $\mu'\supset \mu$ in $L$. In other words, $\mu$ is the maximum element among the elements in $U_{s-1}$ in $L$. An element $\mu$ of $U_{s-1}$ is a \emph{pivot} element of $\mathcal{L}$ if $\mu$ is pivot element of some linear order $L\in \mathcal{L}$.
	
	\begin{lemma}\label{lem:posdim}
		The poset dimension of $P(r,s-1;s)$ is at least $s-r+1$.  
	\end{lemma}
	\begin{proof}
		For the sake of contradiction, suppose that $dim(P(r,s-1;s))\leq s-r$. In other words, $P(r,s-1;s)$ has a realizer $\mathcal{L}$ of size at most $s-r$ and hence the set $Z$ consisting of all pivot elements of $\mathcal{L}$ has at most $s-r$ elements. Label the members of $U_{s-1}$ as $\mu_1,\mu_2,\ldots,\mu_s$ such that $\mu_i=[s]\setminus \{i\}$ for each for each $i\in [s]$. Since $|Z| \leq s-r$, it follows that $U_{s-1}\setminus Z$ has at least ${s\choose s-1} - (s-r)=r$ elements. So we can let $Z'=\{\mu_{i_1},\mu_{i_2},\ldots,\mu_{i_r}\}$ be an $r$-element subset of $U_{s-1}\setminus Z$. Put $\tau=\{i_1,i_2,\ldots,i_r\}$. Observe that there is a linear order $L\in \mathcal{L}$ such that $\tau \supset \mu_{i_1}$ in $L$ since $\tau \not\subset \mu_{i_1}$ as $i_1$ is a member of $\tau$ but not $\mu_{i_1}$. Let $\mu_c$ be the pivot element of $L$. Then $\tau \supset \mu_{c}$ and so $\tau \not\subset \mu_{c}$. Thus $\tau$ must contain $c$ and so $\mu_c=\mu_{i_j}$ for some $1\leq j \leq r$. This contradiction to the fact that $Z'$ contains no pivot element implies the lemma. \qed
	\end{proof}

	\subsection{Lower bound on bend number}\label{subsec-lower-bend}
	For $n,k\in \mathbb{N}$ with $n>k$, define a family of graph $\mathcal{H}_{n,k}$ as follows. Let $Q$ denote the set of $k$-element subsets of $[n]$. A graph $G$ is in $\mathcal{H}_{n,k}$ if its vertex set $V(G)=[n] \cup Q$ and  its edge set $E(G)$ can be partitioned into two sets $E_1$ and $E_2$ such that $E_1 =  \{uv\colon u\neq v,\forall u,v\in [n]\}\cup \{wS\colon \forall S\in Q, \forall w\in S\}$ and $E_2 \subseteq \{ST\colon S\neq T, \forall S,T\in Q\}$. Let $G$ be a graph  that belongs to the family $\mathcal{H}_{n,k}$ for some $n,k\in \mathbb{N}$ with $n>k$. Notice that we do not specify anything about the subgraph induced on vertex set $Q$.
	
	Let $t\in \mathbb{N}$ be a fixed constant. Assume there is a collection $\mathcal{R}$ of $t$-bend paths which gives a $B_t$-VPG representation of $G$. Let $\mathcal{R}_A$ be the collection of paths in $\mathcal{R}$ corresponding to the vertices in $[n]$. In other words, $\mathcal{R}_A=\{P(a)\colon a\in [n], P(a)\in \mathcal{R}\}$. The set of horizontal and vertical lines obtained by extending the horizontal and vertical segments of all the paths in $\mathcal{R}_A$,  along with the set of horizontal and vertical lines passing through both the endpoints of each path in  $\mathcal{R}_A$ is called the \emph{grid} induced by $\mathcal{R}_A$. The horizontal and vertical lines so formed are called \emph{grid lines.} A \emph{vertical strip} is the part of the plane between two consecutive vertical grid lines in the grid induced by $\mathcal{R}_A$. Similarly, a \emph{horizontal strip} is the part of the plane between two consecutive horizontal lines of the grid induced by $\mathcal{R}_A$. By \emph{grid line segments} we mean the part of vertical grid lines in horizontal strips and the part of horizontal grid lines in vertical strips.

	\subsubsection{$k$-sets and good $k$-sets}
\begin{figure}
		\centering
		\begin{tabular}{c}
			\includegraphics[scale=0.55]{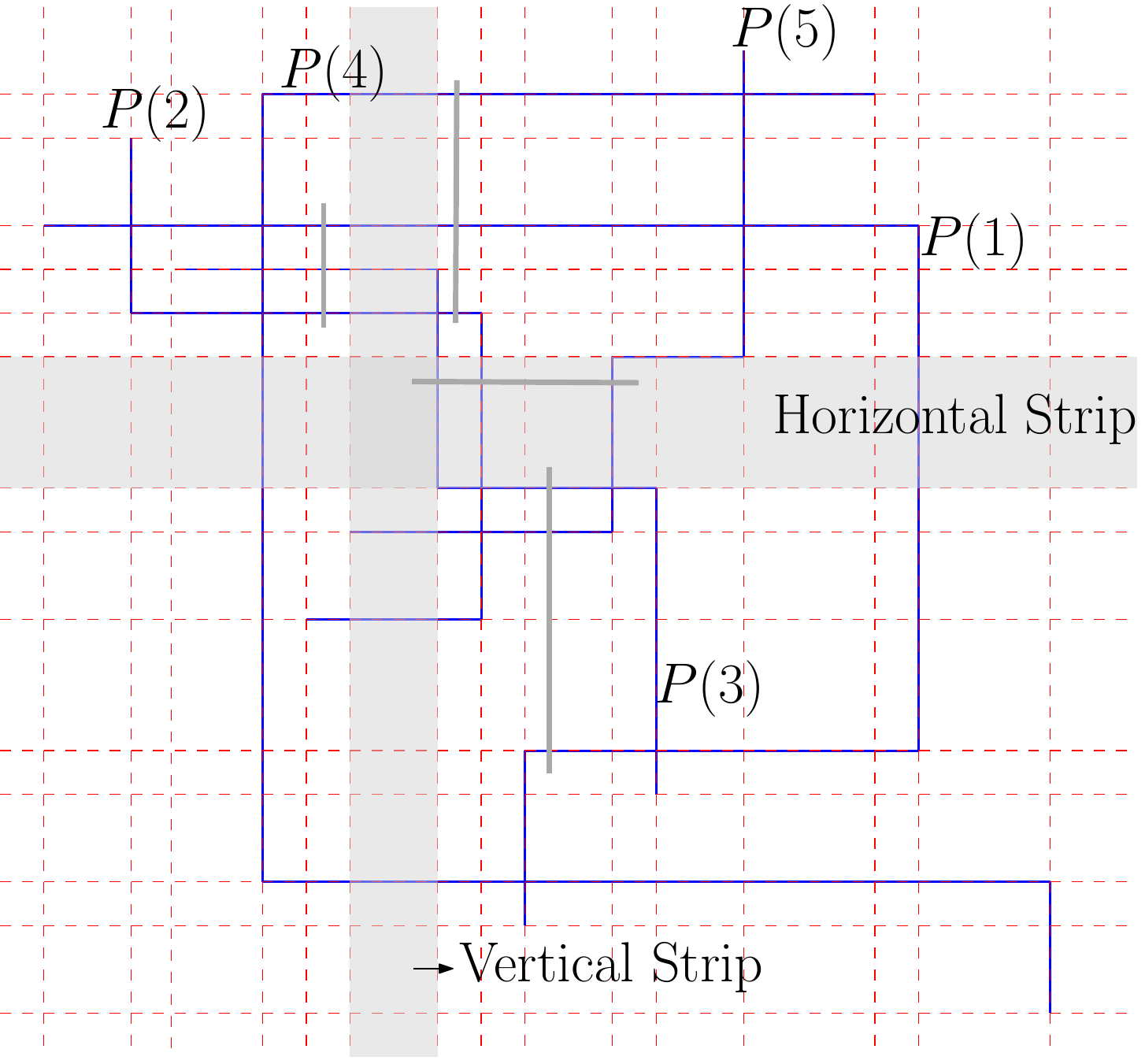}\\
			
		\end{tabular}
		\caption{Grid induced by $\mathcal{R}_A$ of $G\in \mathcal{H}_{5,3}$: (1) dotted lines represent the grid lines, (2) normal lines represent the paths $P(a), a\in[5]$ and (3) thick gray lines intersect some good ${\color{blue} 3}$-sets.}
		\label{fig:grid}
	\end{figure}

	A \emph{$k$-set} is a $k$-element subset of $[n]$. The \emph{distance} between two $k$-sets $S_1,S_2$ is defined as $|S_1\setminus S_2|$. This means if the distance between two $k$-sets $S_1$ and $S_2$ is $d$, then $\mid S_1\cap S_2\mid =k-d$. In particular, when we say the distance between two $k$-sets $S_1$ and $S_2$ is greater than $k-i$, then $|S_1\cap S_2|<i$. Let $S=\{a_1,a_2,\ldots,a_k\}$ be a $k$-set. We say $S$ to be a \emph{vertical good $k$-set} in $\mathcal{R}_A$ if there exists a vertical line segment that intersects  only the paths $P(a_1),P(a_2),\ldots,P(a_k)$. Similarly, $S$ is a \emph{horizontal good $k$-set} in $\mathcal{R}_A$ if there exists a horizontal line segment that intersects only the paths $P(a_1),P(a_2),\ldots,P(a_k)$. A \emph{good $k$-set} is either a \emph{vertical good $k$-set} or a \emph{horizontal good $k$-set} in $\mathcal{R}_A$. See Fig. \ref{fig:grid} for an illustration.
	
	For a fixed constant $t\in\mathbb{N}$, fix $k=2t+16$ and $n=2k^2k!+3$ and let $G_t$ denote any graph in the family $\mathcal{H}_{n,k}$ (the need for such values of $n,k$ will become clear in Observation~\ref{obs:goodkset}). We shall prove that $bend(G_t) > t$. For sake of contradiction, assume $G_t$ has a $B_t$-VPG representation $\mathcal{R}$. Let $\mathcal{R}_A=\{P(a)\colon a\in [n], P(a)\in \mathcal{R}\}$. In the following observation, we  (loosely) upper bound the number of good $k$-sets in $\mathcal{R}_A$.
	
	
	\begin{observation}\label{obs:numgoodset}
		There are at most $2n^2k^2$ good $k$-sets in $\mathcal{R}_A$.
	\end{observation}
	
	\begin{proof}
		We first count the vertical good $k$-sets in $\mathcal{R}_A$. In a vertical strip, there can be at most $n(t+1)$ horizontal segments. Hence, there can be at most $n(t+1)$ vertical good $k$-sets formed by the paths intersecting one vertical strip. Again, there can be at most $n(t+1)$ horizontal segments and $n(t+1)$ vertical segments that intersect a boundary of a vertical strip, resulting in at most $2n(t+1)$ vertical good $k$-sets formed by the paths intersecting a boundary of vertical strip. Since, there are at most $n(t+1)$ vertical strips (and hence at most $n(t+1)+1$ boundaries), there can be at most $4n^2(t+1)^2$ vertical good $k$-sets. Similarly, there can be at most $4n^2(t+1)^2$ horizontal good $k$-sets. Hence, there can be at most $8n^2(t+1)^2$ good $k$-sets.  Since $t=\frac{k-16}{2}$, the result follows. \qed
	\end{proof}
	\begin{remark}
		By more careful counting, we can reduce the number of good $k$-sets in $\mathcal{R}_A$ to at most $n^2k^2$. However, asymptotically, this would not affect our results.
	\end{remark}

	Next in Observation \ref{obs:goodkset}, we prove the existence of a $k$-set which is at a distance greater than $k-3$ away from any good $k$-set. For this we count the number of $k$-sets which are within distance ${k-3}$ from a good $k$-set. Using Observation \ref{obs:numgoodset}, we find an upper bound to the number of $k$-sets which are within distance $k-3$ from any good $k$-set. For value of $n$ as fixed above ($n=2k^2k!+3$), this turns out to be strictly less than $n\choose k$, thereby proving the result.
	
	\begin{observation}\label{obs:goodkset}
		There exists a $k$-set whose distance is greater than $k-3$ from every good $k$-set in $\mathcal{R}_A$.
	\end{observation}
	
	\begin{proof}
		Let $\mathcal{S}$ be the set of all good $k$-sets in $\mathcal{R}_A$. For a good $k$-set $S\in \mathcal{S}$, let $\mathcal{T}_S$ be the set of $k$-sets at a distance $i$ from $S$, where $1 \leq i \leq k-3 $.  The total number of $k$-sets is $n\choose k$. Therefore to prove the observation, it is sufficient to prove the following claim.
		
		\begin{claim}[A]
			\begin{equation*}
			\left|\bigcup\limits_{S\in \mathcal{S}} {\mathcal{T}}_S\right| < {n \choose k}.
			\end{equation*}
		\end{claim}

		\noindent \textit{Proof.} Let $S$ be a good $k$-set in $\mathcal{S}$. Then in order to get a $k$-set that is at a distance $i$ from $S$, we replace $i$ elements in $S$ with $i$ elements not in $S$. So
		\begin{equation*}\label{eq0}
		|\mathcal{T}_S|=\sum_{i=1}^{k-3} {k \choose i} {n-k \choose i}.
		\end{equation*}
		
		In the following inequality, we give an upper bound on $|\mathcal{T}_S|$.

		\begin{equation}\label{eq1}
		|\mathcal{T}_S|=\sum_{i=1}^{k-3} {k \choose i} {n-k \choose i} < (k-3){k \choose \lceil k/2 \rceil}{n-k \choose k-3}
		\end{equation}
		The inequality follows as $n$ is much larger compared to $k$. By Observation~\ref{obs:numgoodset}, total number of good $k$-sets is upper bounded by $2n^2k^2$. Thus using inequality~\ref{eq1}, we get the following upper bound on $\left|\bigcup\limits_{S\in \mathcal{S}} {\mathcal{T}}_S\right|$ i.e. the total number of $k$-sets having distance at most $k-3$ from some good $k$-set in $\mathcal{S}$.
		$$\left|\bigcup\limits_{S\in \mathcal{S}} {\mathcal{T}}_S\right| \leq 2n^2k^2(k-3){k \choose \lceil k/2 \rceil}{n-k \choose k-3}$$
		
		\noindent We require the following two inequalities.
		\begin{subclaim}\label{sl:k!}
			$\text{For }k\geq 16~\text{we have, } k! < \left\lceil \frac{k}{2} \right\rceil! \left\lfloor\frac{k}{2}\right\rfloor ! (k-5)!$
		\end{subclaim}
		
		\begin{subclaim}\label{sl:n!}
			
			$\text{For }n>2(k^2k!+1)~\&~k\geq 16~\text{we have},\hspace{5pt} 2n^2k^2 (n-k)! < \frac{n! (n-2k+3)!}{k! (n-k)!}$
			
		\end{subclaim}
		Before we prove these subclaims, we show how they imply Claim (A).

		\begin{equation*}
		\begin{array}{r@{}l}
		\left|\bigcup\limits_{S\in \mathcal{S}} {\mathcal{T}}_S\right| &{}\leq 2n^2k^2(k-3){k \choose \lceil k/2 \rceil}{n-k \choose k-3}\\
		&{}= 2n^2k^2 (k-3) \frac{k!}{\lceil\frac{k}{2}\rceil! \lfloor\frac{k}{2}\rfloor !} \frac{(n-k)!}{(n-2k+3)!(k-3)!}\\
		&{}= [2n^2k^2 (n-k)!] \left[\frac{k!}{\lceil\frac{k}{2}\rceil! \lfloor\frac{k}{2}\rfloor ! (k-5)!}\right]  \frac{(k-3)}{(k-3)(k-4)(n-2k+3)!} \\
		&{}< \frac{n! (n-2k+3)!}{k! (n-k)!} \frac{1}{(k-4)(n-2k+3)!} \hspace{80pt} \text{using Subclaims~\ref{sl:k!} and~\ref{sl:n!}}\\
		&{}=\binom{n}{k} \frac{1}{(k-4)}\\
		&{}<\binom{n}{k}\hfill \text{since }k\geq 16
		\end{array}
		\end{equation*}    
		So subclaims \ref{sl:k!} and \ref{sl:n!} imply Claim (A).    Now we prove subclaims \ref{sl:k!} and \ref{sl:n!}.
		
		\subsubsection*{Proof of Subclaim \ref{sl:k!}}
		We consider both the cases depending on the parity of $k$ i.e. whether $k=2m$ or $k=2m+1$. In both these cases the required inequality is equivalent to proving $(2m)! < m!m!(2m-5)!$, for $m \geq 8$.
		Since $(2m)! < (2m)^5(2m-5)!$ and $(\frac{m}{3})^m (\frac{m}{3})^m \leq m! m!$ (as for any $n\geq 6$, $\left(\frac{n}{3}\right)^n \leq n! \leq \left(\frac{n}{2}\right)^n$), it is enough to prove the following.
		\begin{align*}    
		&(2m)^5 < \left(\frac{m}{3}\right)^{2m}\\
		\Rightarrow &m > 3^{\frac{2m}{2m-5}}.32^{\frac{1}{2m-5}}
		\end{align*}
		
		For $m=8$, the above inequality holds. Observe, for $m\geq 8$, $3^{\frac{2m}{2m-5}}.32^{\frac{1}{2m-5}}$ is a decreasing function in $m$. Hence for $m \geq 8$ the inequality holds.
		
		Hence $(2m)! < m!\,m!\,(2m-5)!$. This ends the proof of Subclaim~\ref{sl:k!}.

		\subsubsection*{Proof of Subclaim \ref{sl:n!}}
		It is sufficient to prove the following simplified inequality.
		\begin{align*}
		& 2n^2k^2k!\,(n-k)(n-k-1)\ldots (n-2k+4) < n(n-1)(n-2)(n-3)\ldots (n-k+1)\\
		\end{align*}
		Observe that $(n-k)(n-k-1)\ldots (n-2k+4) < (n-3)(n-4)\ldots (n-k+1)$, since every term in RHS is greater than the corresponding term in LHS (as $k\geq 16$).
		
		Hence it is sufficient to prove the following:
		\begin{align*}
		2n^2k^2k!\, &< n(n-1)(n-2)\\
		\Leftrightarrow 2k^2k!\, &< n-3+\frac{2}{n}\\
		\end{align*}
		For $n>2k^2k!+2\,$, the above inequality is satisfied. This ends the proof of Subclaim~\ref{sl:n!}.\\

		\noindent This completes the proof of Observation \ref{obs:goodkset}. \qed
	\end{proof}
	
	Now we find a lower bound for the bend number of $G_t$.
	\subsubsection{Lower Bounding Lemma}
	
	In Observation \ref{obs:numgoodset}, we counted at most $n(t+1)$ good $k$-sets in each horizontal or vertical strip. In $\mathcal{R}_A$, it might be possible that some horizontal or vertical strips do not contain a good $k$-set. Then each such horizontal (resp. vertical) strip intersects with vertical (resp. horizontal) segments of $k' (<k)$ different paths and hence have exactly one good $k'$-set (note that the value of $k'$ might vary among such horizontal or vertical strips).  In the next result, we prove the existence of a $k$-set which has at most two elements common with any good $k$-set and any good $k'$-set.


	\sloppy Suppose $h$ horizontal strips and $v$ vertical strips do not have a good $k$-set. Then they intersect with segments of (say) $k_1, k_2,\ldots, k_h,k_{h+1},\ldots,k_{h+v}$ different paths respectively, with $k_i<k$, for each $i\in[h+v]$. So they have exactly one good $k_i$-set, say $T_i$, for each $i\in[h+v]$. The following lemma proves the existence of a $k$-set which has at most two elements common with any good $k$-set and any good $k'$-set. 
	
	\begin{lemma}\label{lem:lowerbound-0}
		There exists a $k$-set $T$ such that $|T\cap T_i|\leq 2$, for each $i\in[h+v]$; and $|T\cap T_g|\leq 2$, for any good $k$-set $T_g$.
	\end{lemma}

	\begin{proof}
		Suppose wlog a horizontal strip contains no good $k$-set. Then it has vertical segments from $k'<k$ different paths. So it has exactly one good $k'$-set.
		However in Observation \ref{obs:numgoodset}, we accounted for $n(t+1)$ good $k$-sets in this horizontal strip. Now we have none. We form a (dummy) good $k$-set from the $k'$-set present in this horizontal strip by adding $k-k'$ elements.
		
		To visualize this as a good $k$-set, consider $k-k'$ vertical grid lines corresponding to the $k-k'$ elements. Now we have a total of $k$ vertical grid line segments in the horizontal strip (containing the $k'$ vertical segments), which can be intersected by a horizontal line segment, and hence can be considered as a good $k$-set.
		
		We do this for all the $h$ horizontal strips and $v$ vertical strips.
		
		In Observation \ref{obs:goodkset}, we prove that the number of $k$-sets that are within distance $k-3$ from any good $k$-set is strictly less than $n \choose k$ (taking the number of good $k$-sets as $2n^2k^2$). This proves the existence of a $k$-set that is at a distance greater than $k-3$ from every good $k$-set. For the horizontal or vertical strips without a good $k$-set, we have extended the good $k_i$-sets, $T_i$, present in them to exactly one (dummy) good $k$-set in each of them. Now the total number of good $k$-sets is lesser than the number of good $k$-sets considered in the proof of Observation \ref{obs:goodkset} (i.e. $2n^2k^2$). So the number of $k$-sets that are within distance $k-3$ from any good $k$-set is also lesser than the case when there were $2n^2k^2$ good $k$-sets. Hence there exists a  $k$-set $T$ that is at a distance greater than $k-3$ from every good $k$-set (including the dummy ones). This $k$-set has at most two elements common with the good $k_i$-sets and the ``original" good $k$-sets; hence proving the lemma.  \qed
	\end{proof}
	
	By the construction of the graph $G_t$, there exists a vertex $B \in V(G_t)$, such that $N(B)=T$ (this $T$ is the same $k$-set whose existence we proved in Lemma \ref{lem:lowerbound-0}).  We have the following lemma.
	
	\begin{lemma}\label{lem:lowerbound-1}
		$P(B)$ has at least $\frac{k}{2}-1$ bends.
	\end{lemma}

	\begin{proof}
		By Lemma \ref{lem:lowerbound-0}, for any good $k$-set $T_g$,  $|T\cap T_g|\leq  2$.
		Suppose $h$ horizontal strips and $v$ vertical strips intersect with  segments of $k_i(<k)$ different paths, for $i\in[h+v]$, i.e. they each have exactly one good $k_i$-set, say $T_i$.
		By Lemma \ref{lem:lowerbound-0},  $|T\cap T_i|\leq 2$ for any good $k_i$-set $T_i$, for $i\in[h+v]$. This implies that any segment of $P(B)$ in a horizontal or a vertical strip intersects at most two paths in $\mathcal{P}(T_i)$ or $\mathcal{P}(T_g)$. Since $\mid N(B) \mid=k$, $P(B)$ must have at least $\frac{k}{2}$ segments. Hence $P(B)$ has at least $\frac{k}{2}-1$ bends. \qed
	\end{proof}
	\subsection{Completion of Proof of Theorem~\ref{thm:cocomparability}}\label{subsec-combine}
	
	For two fixed constants $t,m\in \mathbb{N}$, fix $k=2t+16$ and $n=\max\{2k^2k!+3,m-3+3k\}$. Define $G_{t,m}$  to be the cocomparability graph obtained by taking the complement of the comparability graph of $P(k,n-1;n)$. The vertex set of $G_{t,m}$ is $A\cup Q$ where $A = \{ a_i \mid i \in [n]  \}$ and $Q=\{b_j \mid j\in \binom{n}{k}\}$. For $j\in \binom{n}{k}$, $b_j$ is indexed by a distinct $k$ element subset of $[n]$. The edge set $E(G_{t,m}) = \{a_ia_j \colon i\neq j;a_i, a_j \in A\} \cup \{ b_ib_j \colon i\neq j; b_i, b_j \in Q \} \cup \{a_ib_j \colon i \text{ belongs to the $k$-element subset of $[n]$ that indexes $b_j$}\}$. Notice that the graph $G_{t,m}$ belongs to the graph family $\mathcal{H}_{n,k}$ as defined in Section~\ref{subsec-lower-bend}. By using similar arguments as used in Section~\ref{subsec-lower-bend}, we can show that $bend(G_{t,m})>t$. Following lemma gives an upper bound on the bend number of $G_{t,m}$. See Section~\ref{sec:lem:bendub} for the proof where we give a $B_{4t+29}$-VPG representation of $G_{t,m}$ i.e $bend(G_{t,m})\leq 4t+29=2k-3$.
	
	\begin{lemma}\label{lem:bendub}
		The bend number of $G_{t,m}$ is at most $4t+29$.
	\end{lemma}
	
	Now using Lemma~\ref{lem:posdim} and Lemma~\ref{lem:bendub}, we conclude that $$dim(G_{t,m}) - bend(G_{t,m}) \geq (n-k+1) - (2k-3) \geq n-3k+4 \geq m-3+3k-3k+4 > m.$$
	The above inequality follows as $n=\max\{2k^2k!+3,m-3+3k\}$. This concludes the proof of Theorem~\ref{thm:cocomparability}.

	\section{Proof of Theorems~\ref{thm:split} and \ref{thm:holefree}}
	
	We shall use the tools developed in Section~\ref{subsec-lower-bend} to prove Theorems \ref{thm:split} and \ref{thm:holefree}.
	
	\subsection{Proof of Theorem \ref{thm:split}}\label{sec:proof-2}

	For $n,k\in \mathbb{N}$ with $n>k$, let $K^k_n$ be the split graph whose vertex set can be partitioned into a clique $C$ having $n$ vertices and an independent set $I$ of size $n\choose k$ such that for any subset $C'\subset C$ with $|C'|=k$, there is a unique vertex $u\in I$ with $N(u)=C'$. For a fixed constant $t\in\mathbb{N}$, fix $k=2t+16$ and $n=2k^2k!+3$ and let $G_t$ denote the graph $K^k_n$. Notice that $G_t$ is a graph in the family $\mathcal{H}_{n,k}$ as defined in Section~\ref{subsec-lower-bend}. Suppose $G_t$ has a $B_t$-VPG representation $\mathcal{R}$.
	Let $\mathcal{R}_A=\{P(a)\colon a\in [n],P(a)\in \mathcal{R}\}$. Then by Lemma~\ref{obs:goodkset}, there is a $k$-set $T$ which has at most two elements common with any good $k'$-set in $\mathcal{R}_A$, where $k'\leq k$.  Since $T$ is a $k$-element subset of $[n]$ there is a vertex $B \in V(G_t)$ such that $N(B)=T$. By Lemma~\ref{lem:lowerbound-1}, $P(B)$ has at least $\frac{k}{2}-1$ bends in $\mathcal{R}$. This contradicts our assumption and hence $bend(G_t)>t$. So for each $t\in \mathbb{N}$, there is a split graph $G$ which have no $B_t$-VPG representation but has $B_{t'}$-VPG representation for some finite $t'> t$ (as split graphs are known to be string graphs; also see Theorem \ref{thm:uperbound-split}). Hence for each $t\in \mathbb{N}$, there exists a $t'>t$ such that $B_t$-VPG-split $\subsetneq B_{t'}$-VPG-split. So there must exist an $m$ with $t\leq m<t'$ such that $B_m$-VPG-split $\subsetneq B_{m+1}$-VPG-split.  This concludes the proof of Theorem~\ref{thm:split}. \qed
	
	\begin{remark}
		In the above discussion, we have a lower bound on the bend number of $K^n_k$ for each $k\in \mathbb{N}$ and sufficiently large $n\in \mathbb{N}$. We can show that bend number of such split graphs is at most $2{n-1\choose k-1}-1$. In Section~\ref{sec:upber-bound}, we prove an  upper bound on the bend number of all split graphs (Theorem~\ref{thm:uperbound-split}).
	\end{remark}

	\subsection{Proof of Theorem~\ref{thm:holefree}}\label{sec:proof-3}
	
	For a fixed constant $t\in \mathbb{N}$, fix $k=2t+16$ and $n=2k^2k!+3$. Let $Q$ denote the set of $k$-element subsets of $[n]$. Let $G_t$ be the graph with vertex set $V(G)=[n] \cup Q$ and edge set $E(G) = \{uv\colon u\neq v,u,v\in [n]\}\cup \{wS\colon S\in Q, w\in S\} \cup \{ST\colon S\neq T,S,T\in Q\}$.
	
	\begin{observation}
		For any $t\in \mathbb{N}$, the graph $G_t$ is in $Forb(C_{\geq 5})$.
	\end{observation}
	
	\begin{proof}
		Assume for the sake of contradiction that for some $t\in \mathbb{N}$ the graph $G_t\notin Forb(C_{\geq 5})$. Then $G_t$ contains an induced cycle $C$ of length at least 5. There exist three vertices $u,v,w\in V(C)$ such that either $\{u,v,w\}\in [n]$ or $\{u,v,w\}\in Q$ (by construction of $G_t$). In either case $u,v,w$ forms a triangle which contradicts our assumption. \qed
	\end{proof}
	
	Notice that the graph $G_t$ belongs to the graph family $\mathcal{H}_{n,k}$ as defined in Section~\ref{subsec-lower-bend}. Hence, by similar arguments used in proof of Theorem~\ref{thm:split}, we can conclude that $t < bend(G_t) \leq 4t+29 t$ and also the proof of Theorem~\ref{thm:holefree}.

	%
	

	\section{Proof of Theorem~\ref{thm:split-2}}\label{sec:proof-4}
	We prove the theorem by showing that for $n\geq 3$, we have $\frac{n-37}{18} \leq bend_p(K^3_n)\leq 2n+4$. Recall that the graph $K^3_n$ is the split graph whose vertex set can be partitioned into a clique $C$ having $n$ vertices and an independent set $I$ of size $n\choose 3$ such that for any subset $C'\subseteq C$ with $|C'|=3$, there is a unique vertex $u\in I$ such that $N(u)=C'$. We use the notion of \emph{contraction} of an edge in our proof. By \emph{contraction} of an edge $uv$ in a graph $G$ we mean  replacement of $u$ and $v$ by a new vertex $w$ adjacent to all the vertices in $N(u)\cup N(v)$.
	
	
	
	First we prove the lower bound of the proper bend number of $K^3_n$. Let $n\geq 3$ be a fixed integer. Let $G$ be a graph isomorphic to $K^3_n$ and $A_n,B_n$ denote the vertices of the clique and independent set of $G$ respectively. Let $\mathcal{R}$ be a proper $B_k$-VPG representation of $G$. For every $b \in B_n$ with $N(b) = \{a_i,a_j,a_l\}$, at least two of their paths, say $P(a_i)$ and $P(a_l)$, intersect $P(b)$ exactly once. Otherwise, we modify $P(b)$ for every $b\in B_n$ as follows. We traverse along the path $P(b)$ from its one end point to another. Consider the sequence of $a$'s whose corresponding $P(a)$'s are being intersected by $P(b)$.
	Call the elements appearing in the first and last position of the sequence as \emph{leaf}. Go on removing the leaf from one endpoint of the sequence if the element corresponding to it is also present elsewhere in the sequence. Repeat this process on the remaining sequence from the other endpoint. When this process stops, both the leaves are distinct. Let $P'(b)$ be the subpath of $P(b)$ that intersects the $P(a)$'s corresponding to the $a$'s in the final sequence. Update $P(b)$ as $P'(b)$.


	So, without loss of generality, $P(b)$ intersects either a horizontal or a vertical segment of $P(a_i)$ and $P(a_l)$ exactly once. Moreover $P(b)$ possibly intersects $P(a_j)$ more than once. For rest of this section, we assume that $P(b)$ intersects either a horizontal or a vertical segment of $P(a_i)$ and $P(a_l)$ exactly once.
	

	\sloppy We form two sets, $S_H$ and $S_V$, from $B_n$ as follows. A vertex $b\in B_n$, with ${N(b) = \{a_i,a_j,a_l\}}$, is in the set $S_H$ if $P(b)$ intersects horizontal segments of at least two of $\{P(a_i),P(a_j),P(a_l)\}$. Similarly, $b \in S_V$ if $P(b)$ intersects vertical segments of at least two of $\{P(a_i),P(a_j),P(a_l)\}$.
	
	\begin{observation}\label{obs:partition}
		The set $B_n$ is union of $S_H$ and $S_V$.
	\end{observation}
	\begin{proof}
		Consider $b \in B_n$ such that $b \notin S_H$. So $P(b)$ intersects horizontal segment of at most one of the paths in $\{P(a_i),P(a_j),P(a_l)\}$. This implies $P(b)$ intersects vertical segments of the other two paths in $\{P(a_i),P(a_j),P(a_l)\}$. Hence $b\in S_V$.  
	\end{proof}
	
	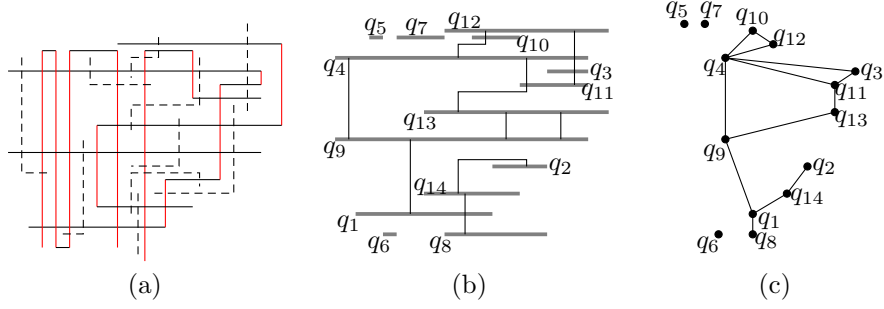
\begin{figure}
		\centering
		\begin{tabular}{ccc}
			\begin{tikzpicture}[scale=0.09]
			%
			
			\draw[solid] (3,5)--(23,5);
			\draw[solid,color=red] (23,5)--(23,12);
			\draw[solid] (23,12)--(31,12);
			\draw[solid,color=red] (31,12)--(31,26);
			\draw[solid] (31,26)--(37,26);
			\draw[solid,color=red] (37,26)--(37,28);        
			\draw[solid] (37,28)--(0,28);

			
			\draw[solid,color=red] (5,2)--(5,31);        
			\draw[solid] (5,31)--(7,31);
			\draw[solid,color=red] (7,31)--(7,2);        
			\draw[solid] (7,2)--(9,2);
			\draw[solid,color=red] (9,2)--(9,31);        
			\draw[solid] (9,31)--(16,31);
			\draw[solid,color=red] (16,31)--(16,2);        
			
			\draw[solid] (0,16)--(37,16);
			
			\draw[solid,color=red] (20,0)--(20,31);        
			\draw[solid] (20,31)--(27,31);
			\draw[solid,color=red] (27,31)--(27,24);        
			\draw[solid] (27,24)--(37,24);
			
			\draw[solid] (16,32)--(40,32);
			\draw[solid,color=red] (40,32)--(40,20);        
			\draw[solid] (40,20)--(13,20);
			\draw[solid,color=red] (13,20)--(13,8);        
			\draw[solid] (13,8)--(27,8);

			\draw[densely dashed] (35,35)--(35,22);
			\draw[densely dashed] (33,23)--(33,10)--(21,10);
			\draw[densely dashed] (25,21)--(25,14)--(18,14);
			\draw[densely dashed] (28,30)--(28,23)--(18,23)--(18,19);
			\draw[densely dashed] (12,30)--(12,26)--(21,26);
			\draw[densely dashed] (2,30)--(2,13)--(6,13);
			\draw[densely dashed] (19,1)--(19,10);
			\draw[densely dashed] (18,7)--(18,13)--(28,13)--(28,11);
			\draw[densely dashed] (8,4)--(11,4)--(11,18);
			\draw[densely dashed] (22,33)--(22,30)--(18,30)--(18,27);
			\end{tikzpicture}
			&\begin{tikzpicture}[scale=0.09]
			%
			
			\draw[solid, color=gray, line width=0.5mm] (3,5)--(23,5);
			\draw[solid, color=gray, line width=0.5mm] (23,12)--(31,12);
			\draw[solid, color=gray, line width=0.5mm] (31,26)--(37,26);
			\draw[solid, color=gray, line width=0.5mm] (37,28)--(0,28);
			\draw[solid, color=gray, line width=0.5mm] (5,31)--(7,31);
			\draw[solid, color=gray, line width=0.5mm] (7,2)--(9,2);
			\draw[solid, color=gray, line width=0.5mm] (9,31)--(16,31);
			\draw[solid, color=gray, line width=0.5mm] (16,2)--(31,2);
			\draw[solid, color=gray, line width=0.5mm] (0,16)--(37,16);
			\draw[solid, color=gray, line width=0.5mm] (20,31)--(27,31);
			\draw[solid, color=gray, line width=0.5mm] (27,24)--(37,24);
			\draw[solid, color=gray, line width=0.5mm] (16,32)--(40,32);
			\draw[solid, color=gray, line width=0.5mm] (40,20)--(13,20);
			\draw[solid, color=gray, line width=0.5mm] (13,8)--(27,8);
			
			\node at (2,3.5) {$q_1$};
			\node at (33,12) {$q_{2}$};
			\node at (39,26) {$q_{3}$};
			\node at (0,26.5) {$q_{4}$};
			\node at (6,32.5) {$q_{5}$};
			\node at (6.5,0.5) {$q_{6}$};
			\node at (12,32.5) {$q_{7}$};
			\node at (15.5,0.5) {$q_{8}$};
			\node at (0,14.5) {$q_{9}$};
			\node at (29,30) {$q_{10}$};
			\node at (38.5,23) {$q_{11}$};
			\node at (19,33.5) {$q_{12}$};
			\node at (12.5,18.5) {$q_{13}$};
			\node at (14,9) {$q_{14}$};

			\draw[solid] (35,32)--(35,24);
			\draw[solid] (33,20)--(33,16);
			\draw[solid] (25,20)--(25,16);
			\draw[solid] (28,28)--(28,24)--(28,23)--(18,23)--(18,20);
			\draw[solid] (2,28)--(2,16);
			\draw[solid] (19,2)--(19,8);
			\draw[solid] (18,8)--(18,13)--(28,13)--(28,12);
			\draw[solid] (11,5)--(11,16);
			\draw[solid] (22,32)--(22,31)--(22,30)--(18,30)--(18,28);
			\end{tikzpicture}&\begin{tikzpicture}[scale=0.09]
			\draw[fill] (20,32) circle (15pt);
			\draw[fill] (23,30) circle (15pt);
			\draw[fill] (16,28) circle (15pt);
			\draw[fill] (35,26) circle (15pt);
			\draw[fill] (32,24) circle (15pt);
			\draw[fill] (32,20) circle (15pt);
			\draw[fill] (16,16) circle (15pt);
			\draw[fill] (28,12) circle (15pt);
			\draw[fill] (25,8) circle (15pt);
			\draw[fill] (20,5) circle (15pt);
			\draw[fill] (20,2) circle (15pt);
			\draw[fill] (15,2) circle (15pt);
			\draw[fill] (13,33) circle (15pt);
			\draw[fill] (10,33) circle (15pt);
			
			\node at (20,34) {$q_{10}$};
			\node at (25.5,31) {$q_{12}$};
			\node at (14.5,26.5) {$q_{4}$};
			\node at (37.5,26) {$q_{3}$};
			\node at (34.5,23) {$q_{11}$};
			\node at (34.5,19) {$q_{13}$};
			\node at (14.5,14) {$q_{9}$};
			\node at (30.5,12) {$q_{2}$};
			\node at (28,7.5) {$q_{14}$};
			\node at (22.5,4) {$q_{1}$};
			\node at (22,1) {$q_{8}$};
			\node at (13.5,0.5) {$q_{6}$};
			\node at (14,35) {$q_{7}$};
			\node at (9,35) {$q_{5}$};
			
			\draw[solid] (20,32)--(23,30)--(16,28)--(20,32);
			\draw[solid] (16,28)--(35,26)--(32,24)--(16,28)--(16,16)--(32,20)--(32,24);
			\draw[solid] (16,16)--(20,5)--(25,8)--(28,12);
			\draw[solid] (20,5)--(20,2);
			\end{tikzpicture}\\
			(a)&(b)&(c)
		\end{tabular}
		\caption{(a) Part of $B_k$-VPG representation of $K_5^3$ where for $a\in A_n$ and $b\in B_n$, $P(a)$'s are normal lines and $P(b)$'s are dashed lines, (b) Vertices of $F_h$ are represented by the thick gray lines and part of subpaths $P(b)$ as normal lines (c) The planar graph $F_h$.}
		\label{fig:planar}
	\end{figure}

	We shall count the cardinality of the sets $S_H$ and $S_V$. Notice that, by Observation~\ref{obs:partition}, $n\choose 3$  $=|B_n|$ $\leq |S_H| + |S_V|$. Let $\mathcal{R}_A=\{P(a)\colon a\in A_n, P(a)\in \mathcal{R}\}$. We define two sets $R_H$ and $R_V$ as collection of all horizontal segments and all vertical segments of all paths in $\mathcal{R}_A$ respectively.
	Now we construct two graphs $F_h$ and $F_v$. The vertex set and edge set of $F_h$ are denoted by $V(F_h)$ and $E(F_h)$ (similarly for $F_v$).  
	The vertex set of $F_h$ has a bijection with $R_H$. For two vertices $u, v \in V(F_h)$, $uv \in E(F_h)$ if for some $b \in B_n$, there is a sub-path of $P(b)$ that intersects $u$ and $v$ and does not intersect any other horizontal segment in $R_H$. (Remove multiple edges if any.) See Fig. \ref{fig:planar} for an illustration.  Similarly the vertex set of $F_v$ has a bijection with $R_V$. For two vertices $u, v \in V(F_v)$, $uv \in E(F_v)$ if for some $b \in B_n$, there is a sub-path of $P(b)$ that intersects $u$ and $v$ and does not intersect any other vertical segment in $R_V$. (Remove multiple edges if any.) Since $B_n$ is an independent set, none of the $P(b)$'s intersect with each other. So none of the edges of $F_h$ (and $F_v$) intersect each other, except at the vertices. Hence $F_h$ and $F_v$ are planar graphs. They each have at most $n(\frac{k}{2}+1)$ vertices (since the number of bends in a path is at most $k$, number of horizontal or vertical segments is at most $\frac{k}{2}+1$). Since a planar graph $G$ has at most $3|V(G)|-6$ edges, each of $F_H$ and $F_V$ will have at most $3n(\frac{k}{2}+1)$ edges.
	
	Let $E_1=\{uv\in E(F_h)\colon$ there is a vertex $a\in A_n$ such that the horizontal segments corresponding to $u$ and $v$ belong to $P(a)\}$. Let $E_2=\{uv\in E(F_v)\colon$ there is a vertex $a\in A_n$ such that the vertical segments corresponding to $u$ and  $v$ belong to $P(a)\}$. Let $F'_h$ be the graph obtained by contracting all the edges in $E_1$ and $F'_v$ be the graph obtained by contracting all the edges in $E_2$. Since the graphs $F'_h$ and $F'_v$ are obtained by contracting edges in planar graphs $F_h$ and $F_v$ respectively, they are also planar. Hence each of $F'_h$ and $F'_v$ have at most $3n(\frac{k}{2}+1)$ edges.
	
	\begin{observation}
		$|S_H| \leq 3n(\frac{k}{2}+1)(n-2)$ and     $|S_V| \leq 3n(\frac{k}{2}+1)(n-2)$.
	\end{observation}
	\begin{proof}
		
		We shall only prove that $\mid S_H \mid \leq 3n(\frac{k}{2}+1)(n-2)$. Consider a vertex $b \in S_H$ with $N(b) = \{a_i, a_j, a_l\}$. By definition, $P(b)$ intersects horizontal segments of at least two of $\{P(a_i),P(a_j),P(a_l)\}$.
		
		\sloppy     Suppose $P(b)$ intersects horizontal segments of exactly, say $P(a_i)$ and $P(a_j)$. Then all intersection points of $P(b)$ and $P(a_l)$ lie on vertical segments of $P(a_l)$. In this case, there is an edge $xy\in E(F'_h)$ such that the horizontal segments corresponding to $x$ and $y$ lies on paths in $P(a_i)$ and $P(a_j)$ respectively.    
		
		Suppose $P(b)$ intersects horizontal segments of all of $\{P(a_i),P(a_j),P(a_l)\}$, then there exists a 2-path $xyz$ in $F'_h$ such that the horizontal segments corresponding to $x,y,z$ are parts of $P(a_i),P(a_j),P(a_l)$ respectively.
		

		
		\sloppy Therefore, for every $b \in S_H$ with $N(b)=\{a_i,a_j,a_l\}$, there is an edge $uv \in E(F'_h)$ such that $u$ corresponds to a horizontal segment of $P(x)\in \{P(a_i),P(a_j),P(a_l)\}$ and $v$ corresponds to a horizontal segment of $P(y)\in \{P(a_i),P(a_j),P(a_l)\}\setminus \{P(x)\}$.
		
		Let $u'v'$ be an edge of $E(F'_h)$, and let $u'$ corresponds to a horizontal segment of $P(x')$ and $v'$ corresponds to a horizontal segment of $P(y')$.  The total number of vertices in the set $\{q\in S_H\colon \{x',y'\}\subset N(q)\}$ is at most $n-2$. So $\mid S_H \mid \leq (n-2)\mid E(F'_h)\mid \leq 3n(\frac{k}{2}+1)(n-2)$.
		
		Similarly we get $\mid S_V \mid \leq 3n(\frac{k}{2}+1)(n-2)$ by repeating the above analysis on $F'_v$.  \qed
	\end{proof}
	
	Therefore, we have the following.
	\begin{align*}
	&{n\choose 3} \, \leq |S_H | + | S_V|\\
	\Rightarrow & \frac{n(n-1)(n-2)}{6}\, \leq 6n(\frac{k}{2}+1)(n-2)\\
	\Rightarrow & \frac{n-37}{18}\, \leq k\\
	\end{align*}
	Therefore $\frac{n-37}{18} \leq bend_p(K^3_n)$ for all $n\geq 3$. In the following lemma we shall show that, $bend_p(K^3_n)\leq 2n+4$ for all $n\geq 3$. See Section~\ref{sec:lem:ubound-2} for the proof where we give a $B_{2n+4}$-VPG representation of $K^3_n$.
	
	\begin{lemma}\label{lem:ubound-2}
		For all $n\geq 3$, $bend_p(K^3_n)\leq 2n+4$.
	\end{lemma}
	
	To complete the proof of Theorem~\ref{thm:split-2}, for given $t\in \mathbb{N}$, fix $n=18t+38$ and let $G_t$ denote the graph $K^3_n$. From the above discussions we can infer that $t<bend_p(G_t)\leq 36t+80$. Therefore for all $t\in \mathbb{N}$, $PB_t$-VPG-split $\subsetneq PB_{36t+80}$-VPG-split.

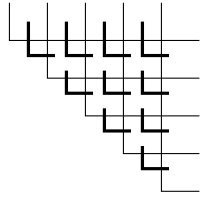
\begin{figure}
		\centering
		\begin{tikzpicture}
		\draw (0,0.5) -- (0,0) -- (2.5,0);
		\draw (0.5,0.5) -- (0.5,-0.5) -- (2.5,-0.5);
		\draw (1,0.5) -- (1,-1) -- (2.5,-1);
		\draw (1.5,0.5) -- (1.5,-1.5) -- (2.5,-1.5);
		\draw (2,0.5) -- (2,-2) -- (2.5,-2);
		\draw[very thick] (0.25,0.25) -- (0.25,-0.2)-- (0.6,-0.2);
		\draw[very thick] (0.75,0.25) -- (0.75,-0.2)-- (1.1,-0.2);
		\draw[very thick] (1.25,0.25) -- (1.25,-0.2)-- (1.6,-0.2);
		\draw[very thick] (1.75,0.25) -- (1.75,-0.2)-- (2.1,-0.2);
		
		\draw[very thick] (0.75,-0.4) -- (0.75,-0.7)-- (1.1,-0.7);
		\draw[very thick] (1.25,-0.4) -- (1.25,-0.7)-- (1.6,-0.7);
		\draw[very thick] (1.75,-0.4) -- (1.75,-0.7)-- (2.1,-0.7);
		
		\draw[very thick] (1.25,-0.9) -- (1.25,-1.2)-- (1.6,-1.2);
		\draw[very thick] (1.75,-0.9) -- (1.75,-1.2)-- (2.1,-1.2);
		
		\draw[very thick] (1.75,-1.4) -- (1.75,-1.7)-- (2.1,-1.7);
		\end{tikzpicture}
		\caption{A $PB_1$-VPG representation of $K^2_5$.}\label{fig:K-2-n}
	\end{figure}

	\begin{remark}
		Asinowski et al.~\cite{asinowski2012} proved that the bend number of $K^3_{33}$ is greater than 1. From Lemma~\ref{lem:ubound-2}, it is clear that $bend(K^3_{33}) \leq bend_p(K^3_{33}) \leq 70$. However, it is not difficult to verify that $bend_p(K^2_n)\leq 1$ for all $n\geq 2$. See Figure~\ref{fig:K-2-n} for a $PB_1$-VPG representation of $K^2_5$. In Section~\ref{sec:upber-bound}, we shall prove an upper bound on the bend number of split graphs in general (Theorem~\ref{thm:uperbound-split}).
	\end{remark}
	
	Proof of the following corollary follows from the proof of Lemma~\ref{lem:ubound-2} (reader may verify that the upper bound for $K^3_n$ given in the next corollary is better than that of Theorem~\ref{thm:uperbound-split}).
	
	\begin{corollary}
		Let $G$ be a split graph with clique partition $C$ and independent set partition $I$. If $\mid N(v) \mid \leq 3$ for all $v\in I$ then $bend_p(G)\leq 2n+4$ where $n=\mid C \mid $.  
	\end{corollary}

	\section{Upper bounds}\label{sec:upber-bound}
	
	In this section, first we prove an upper bound on the bend number of split graphs. Then in Section~\ref{sec:lem:ubound-2} we prove Lemma~\ref{lem:ubound-2} and in Section~\ref{sec:lem:bendub} we prove Lemma~\ref{lem:bendub}.
	
	\begin{theorem}\label{thm:uperbound-split}
		Let $G$ be a split graph with clique partition $C$ and independent set partition $I$. Then $bend(G)\leq 2\Delta_c-1$ where $\Delta_c=\max\{\mid N(v)\cap I\mid \colon v\in C\}$.
	\end{theorem}
	
	\begin{proof}
		Let $C=\{v_1,v_2,\ldots,v_c\}$ and $I=\{u_1,u_2,\ldots,u_m\}$. For each $j\in [m]$, we first draw horizontal segment $l_j$ whose left endpoint is at $(2j,2j)$ and right endpoint is at $(2j+1,2j)$. For each $j\in [m]$, let the label of $l_j$ be $N(u_j)$.

		Now we draw the paths corresponding to the vertices of $C$ and $I$. For each $j\in[m]$, we draw $P(u_j)$ as a vertical segment whose top endpoint is at $(2j+\frac{1}{2},2j-\frac{1}{2})$ and bottom endpoint is at $(2j+\frac{1}{2},2j+\frac{1}{2})$. For each $i\in [c]$, let $L_i=\{l_{i_1}, l_{i_2} ,\ldots,l_{i_t}\}$ with $t=\mid N[v_i]\cap I\mid $ and $i_1<i_2<\ldots<i_t$ be the horizontal segments such that $v_i$ belongs to the label of $l_{i_j}$ for each $j\in [t]$. For each $i\in [c]$, we draw $P(v_i)$ by joining $(0,0)$ with the left endpoint of $l_{i_1}$ using a 1-bend rectilinear path and then for each $j\in [ \mid N(v_i)\cap I \mid - 1]$ we join right endpoint of $l_{i_j}$ with the left endpoint of $l_{i_{j+1}}$ using 1-bend rectilinear path.
		
		It is not difficult to verify that union of $\{P(v)\}_{v\in C}$ and $\{P(u)\}_{u\in I}$ is a valid VPG representation of $G$. The bend number of $P(u)$ for each $u\in I$ is $0$. For each $v\in C$, $P(v)$ has exactly $2(\mid N(v)\cap I \mid-1)+1$ bends i.e. at most $2\Delta_c-1$ bends where $\Delta_c=\max\{\mid N(v)\cap I\mid \colon v\in C\}$. This concludes the proof. \qed
	\end{proof}

	\subsection{Proof of Lemma~\ref{lem:ubound-2}}\label{sec:lem:ubound-2}
	Now we shall prove that $bend_p(K_n^3) \leq 2n+4$, for all $n\geq 3$. For the remainder of this section fix an integer $n\geq 3$. Recall that the clique partition of $K^3_n$ has $n$ vertices and the independent set partition of $K^3_n$ has $n\choose 3$ vertices. Let $V_c=\{a_1,a_2,\ldots,a_n\}$ denote the vertices in the clique partition of $K^3_n$ and $V_I$ represent the independent set partition of $K^3_n$.
	
	To draw the $PB_{2n+4}$-VPG representation of $K^3_n$ we use the following standard graph theory result~\cite{Harary69a}.
	
	\begin{lemma}\label{lm:hamcycle}
		The complete graph on $4s+1$ vertices can be decomposed into $2s$ edge-disjoint Hamiltonian cycles.
	\end{lemma}
	
	An algorithm for the above decomposition can also be found in \cite{Harary69a}. Given $n$ we add at most $3$ dummy vertices such that the sum is of the form $4s+1$, for some $s\in \mathbb{N}$. Let $H$ be the complete graph on union of the vertex set $\{a_1, a_2, \ldots, a_n\}$ and the dummy vertices. Using Lemma \ref{lm:hamcycle} we can infer that there is a set $\mathcal{C}$ consisting of $2s$ edge disjoint hamiltonian cycles that partitions $E(H)$. Let $\{C_1,C_2,\ldots,C_{2s}\}$ be the cycles in $\mathcal{C}$. For each $i\in [2s]$, we form sequence $S_i$ by starting at $a_i$ and including only the indexes of vertices encountered while traversing along the cycle $C_i$ (we do not consider the dummy vertices in the sequence). So $S_i$ begins and ends with $i$ and has other $n-1$ numbers in between. For any pair $(a_i,a_j)$ we have an edge $a_ia_j\in E(H)$ and therefore $a_ia_j$ is an edge in some hamiltonian cycle of $\mathcal{C}$. The following observation directly follows:
	
	\begin{observation}
		For every distinct $i,j \in [n]$, $i$ and $j$ lie consecutively in some sequence.
	\end{observation}
	
	\paragraph{Construction:} Fix origin $O$ with the usual $X$ and $Y$ axis.
	For each odd $i\in [s]$ we fix a square region $s_i$ of dimensions $(n+2)\times(n+2)$ whose bottom left corner and top left corner are fixed at $(2ni,0)$ and $(2ni,n+2)$. Similarly for each even $i\in [s]$ we fix a square region $s_i$ of dimensions $(n+2)\times(n+2)$ whose bottom left corner and top left corner are fixed at $(2ni,0.5)$ and $(2ni,n+2.5)$.
	
	Now for each odd $i\in [s]$, we draw horizontal line segments at $y=1, y=2,\ldots, y=n+1$ such that the horizontal lines partition the square $s_i$. Similalrly for each even $i\in [s]$, we draw horizontal line segments at $y = 1.5,y=2.5,\ldots, y=n+1.5$  such that the horizontal lines partition the square $s_i$. Also for each $i\in [s]$, we draw vertical line segments at $x = 2ni+1, x=2ni+2, \ldots, x=2ni+n+1$ such that the vertical lines partitions the square $s_i$.
	
	For each $i\in [s]$, we use the square region $s_i$ to represent sequences $S_i$ and $S_{s+i}$. Let the $j^{th}$ vertical line from left in the square $s_i$ be labeled as the $j^{th}$ entry in sequence $S_i$. Let the $j^{th}$ horizontal line from bottom in the $i^{th}$ square be labeled as the $j^{th}$ entry in sequence $S_{s+i}$.

	For each $i\in [2s]$, the sequence $S_i$ begins and ends with $i$, so the topmost and bottommost line segments of each square have the same label. Similarly, the leftmost and rightmost line segments of each square region have the same label.
	
\begin{figure}
		\centering
		\begin{tikzpicture}[scale=0.67]
		\draw [solid] (0,0)--(0,2);
		\draw [dashed] (0.5,0)--(0.5,2);
		\draw [dashed] (1,0)--(1,2);
		\draw [dashed] (1.5,0)--(1.5,2);
		\draw [solid] (2,0)--(2,2);
		\draw [solid] (-0.2,1)--(2.2,1);
		\draw [dashed] (-0.2,0.3)--(2.2,0.3);
		\draw [dashed] (-0.2,0.6)--(2.2,0.6);
		\draw [dashed] (-0.2,1.3)--(2.2,1.3);
		\draw [dashed] (-0.2,1.6)--(2.2,1.6);
		
		\node at (-0.1,-0.1) {$l$}; \node at (-0.1,2.1) {$l$}; \node at (2.1,-0.1) {$l$}; \node at (2.1,2.1) {$l$}; \node at (-0.3,0.9) {$l$}; \node at (2.2,0.9) {$l$};
		
		\draw[->,solid] (2.5,1)--(4,1);
		
		\draw [solid] (4,2)--(5,2)--(5,0)--(5.1,0)--(5.1,1)--(6.9,1)--(6.9,2)--(7,2)--(7,0)--(8,0);
		\draw [dashed] (5.5,0)--(5.5,2);
		\draw [dashed] (6,0)--(6,2);
		\draw [dashed] (6.5,0)--(6.5,2);
		\draw [dashed] (4.8,0.3)--(7.2,0.3);
		\draw [dashed] (4.8,0.6)--(7.2,0.6);
		\draw [dashed] (4.8,1.3)--(7.2,1.3);
		\draw [dashed] (4.8,1.6)--(7.2,1.6);
		
		\node at (4.9,-0.1) {$l$}; \node at (5.1,2.1) {$l$}; \node at (6.9,-0.1) {$l$}; \node at (7.1,2.1) {$l$}; \node at (4.9,0.9) {$l$}; \node at (7.1,0.9) {$l$};
		
		\node at (3.2,1.2) {\footnotesize $\text{Oper}^n$ 1};
		
		\end{tikzpicture}
		\hspace{8pt}
		\begin{tikzpicture}[scale=0.67]
		\draw [dashed] (0,0)--(0,2);
		\draw [dashed] (0.5,0)--(0.5,2);
		\draw [solid] (1,0)--(1,2);
		\draw [solid] (-0.1,0.1)--(2.1,0.1);
		\draw [solid] (-0.1,1.9)--(2.1,1.9);
		\draw [dashed] (1.5,0)--(1.5,2);
		\draw [dashed] (2,0)--(2,2);
		\draw [dashed] (-0.2,1)--(2.2,1);
		\draw [dashed] (-0.2,0.3)--(2.2,0.3);
		\draw [dashed] (-0.2,0.6)--(2.2,0.6);
		\draw [dashed] (-0.2,1.3)--(2.2,1.3);
		\draw [dashed] (-0.2,1.6)--(2.2,1.6);
		
		\node at (-0.1,-0.2) {$l$}; \node at (-0.1,2.2) {$l$}; \node at (2.1,-0.2) {$l$}; \node at (2.1,2.2) {$l$}; \node at (1,-0.2) {$l$}; \node at (1,2.2) {$l$};
		
		\draw[->,solid] (2.5,1)--(4,1);
		
		\draw [solid] (4,1.9)--(7,1.9)--(7,1.8)--(6,1.8)--(6,0.2)--(5,0.2)--(5,0.1)--(8,0.1);
		\draw [dashed] (5.5,0)--(5.5,2);
		\draw [dashed] (6.5,0)--(6.5,2);
		\draw [dashed] (4.8,0.3)--(7.2,0.3);
		\draw [dashed] (4.8,0.6)--(7.2,0.6);
		\draw [dashed] (4.8,1)--(7.2,1);
		\draw [dashed] (4.8,1.3)--(7.2,1.3);
		\draw [dashed] (4.8,1.6)--(7.2,1.6);

		\node at (4.9,-0.1) {$l$}; \node at (5.1,2.1) {$l$}; \node at (6.9,-0.1) {$l$}; \node at (7.1,2.1) {$l$}; \node at (6,-0.2) {$l$}; \node at (6,2.2) {$l$};

		\node at (3.2,1.2) {\footnotesize $\text{Oper}^n$ 2};
		
		\end{tikzpicture}
		
		\begin{tikzpicture}[scale=0.7]
		\draw [dashed] (0,0)--(0,2);
		\draw [dashed] (0.5,0)--(0.5,2);
		\draw [solid] (1,0)--(1,2);
		\draw [dashed] (-0.2,0.1)--(2.2,0.1);
		\draw [dashed] (-0.2,1.9)--(2.2,1.9);
		\draw [dashed] (1.5,0)--(1.5,2);
		\draw [dashed] (2,0)--(2,2);
		\draw [dashed] (-0.2,1)--(2.2,1);
		\draw [dashed] (-0.2,0.3)--(2.2,0.3);
		\draw [dashed] (-0.2,0.6)--(2.2,0.6);
		\draw [solid] (-0.2,1.3)--(2.2,1.3);
		\draw [dashed] (-0.2,1.6)--(2.2,1.6);
		
		\node at (0.9,-0.2) {$l$}; \node at (0.9,2.2) {$l$}; \node at (-0.3,1.3) {$l$}; \node at (2.3,1.3) {$l$};
		
		\draw[->,solid] (2.5,0.8)--(4,0.8);
		
		\draw [solid] (4,1.3)--(5.9,1.3)--(5.9,2)--(6,2)--(6,0)--(6.1,0)--(6.1,1.3)--(8,1.3);
		\draw [dashed] (5,0)--(5,2);
		\draw [dashed] (5.5,0)--(5.5,2);
		\draw [dashed] (4.8,0.1)--(7.2,0.1);
		\draw [dashed] (4.8,1.9)--(7.2,1.9);
		\draw [dashed] (6.5,0)--(6.5,2);
		\draw [dashed] (7,0)--(7,2);
		\draw [dashed] (4.8,1)--(7.2,1);
		\draw [dashed] (4.8,0.3)--(7.2,0.3);
		\draw [dashed] (4.8,0.6)--(7.2,0.6);
		\draw [dashed] (4.8,1.6)--(7.2,1.6);
		
		\node at (5.9,-0.2) {$l$}; \node at (6.1,2.2) {$l$}; \node at (4.8,1.5) {$l$}; \node at (7.2,1.5) {$l$};
		
		\node at (3.2,1) {\footnotesize $\text{Oper}^n$ 3};
		
		\end{tikzpicture}
		\caption{Operations for drawing paths.}\label{fig:operation}
	\end{figure}
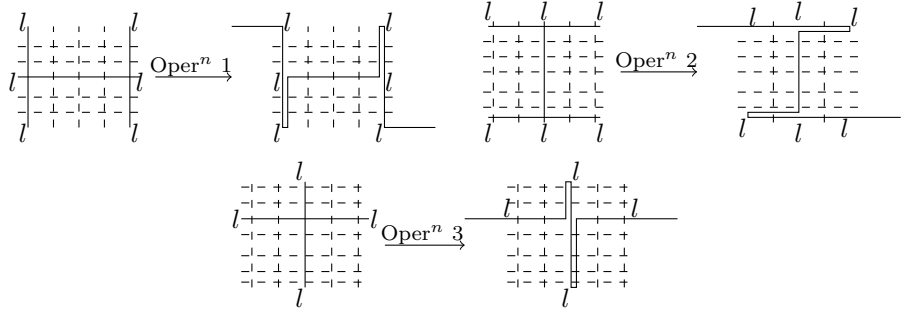

	Now for each $i\in [n]$ we do a few modifications to make sure that the vertical line segment labeled $l$ does not intersect the horizontal line segment labeled $l$; in fact, we merge them into one rectilinear path labeled $l$. This is depicted in the three operations shown in Figure~\ref{fig:operation}.

	Observe that every sequence begins with a distinct number, so horizontal segments labeled $l$ undergoes either operation $1$ or $2$ just once. In rest of the squares, it undergoes operation $3$. The set of rectilinear paths so obtained have the following properties: no two segments overlap, every intersection point is contained in exactly two segments and whenever two segments intersect they cross each other.
	
	Recall that $V_C$ is the clique partition of $K^3_n$. Now for each $l\in [n]$, we can join the rectilinear curves labeled $l$ to form a single path $P(a_l)$ corresponding to the vertex $a_l\in V_c$. Observe that the set of paths $\mathcal{R}_A=\{P(a)\}_{a\in V_c}$ is a $PB_k$-VPG representation of the complete graph induced by $V_c$ in $K^3_n$. We shall count the value of $k$ later.
	
	
	Recall that $V_I$ is the independent set partition of $K^3_n$. Now for each $b\in V_I$ we add paths of $P(b)$ in $\mathcal{R}_A$. Let $N(b)=\{a_p,a_q,a_r\}$, where $1 \leq p<q<r\leq n$. Let $a_{p}a_{q}$ be an edge in cycle $C_i\in \mathcal{C}$. So $p, q$ appear consecutively in $S_i$, and hence $P(a_{p})$ and $P(a_{q})$ appear consecutively in some square region. In the same square region, $P(a_{r})$ intersects both $P(a_{p})$ and $P(a_{q})$ orthogonally. Now consider a small rectangle just enclosing these intersection points only. Let $P(b)$ be any two consecutive sides of this rectangle. Observe that $P(b)$ has only one bend and it intersects only $P(a_{p}), P(a_{q})$ and $P(a_{r})$. Also observe that all the $P(b_i)$'s do not intersect each other. Therefore, $\mathcal{R}=\{P(a)\}_{a\in V_c}\cup \{P(b)\}_{b\in V_I}$ is a $PB_k$-representation of $K^3_n$.
	
	\paragraph{Counting:} Now we shall count the value of $k$. Operation $1$ needs the most number of bends, hence the path, say $P(a_p)$ undergoing this operation will have maximum bend number amongst all $P(a_i)$'s, for $i \in [n]$. There are $s$ square regions and between each consecutive pair, we need $2$ bends. In the one square region where $P(a_p)$ undergoes operation $1$ it needs $8$ bends and in the rest $s-1$ regions it needs $6$ bends. Hence $P(a_p)$ has total $8+6(s-1)+2(s-1)=8s$ bends which is at most $2n+4$ bends.
	
	So we have a $PB_{2n+4}$-VPG representation of $K^3_n$ proving that $bend_p(K^3_n) \leq 2n+4$. This completes the proof of Lemma~\ref{lem:ubound-2}. See below for an example.

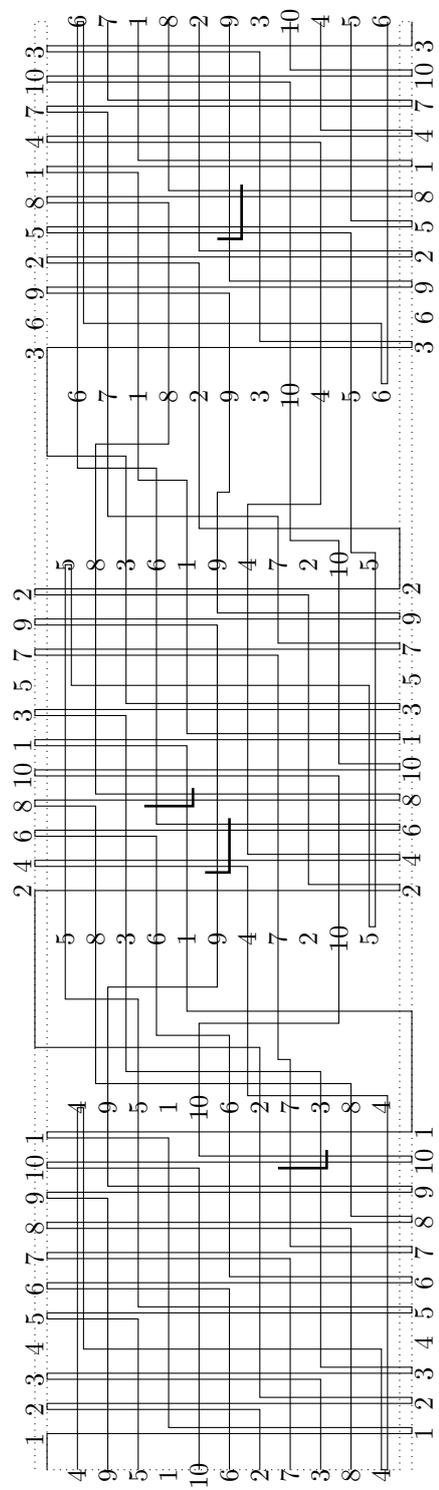
\begin{sidewaysfigure}
		\centering
		\vspace{10cm}
		\hspace{0cm}
		\begin{tikzpicture}[scale=0.8]
		
		\draw[dotted] (0,0)--(24,0)--(24,0.2)--(0,0.2)--(0,6)--(24,6)--(24,6.2)--(0,6.2);

		\draw[solid] (0,6)--(0.6,6)--(0.6,0)--(0.7,0)--(0.7,4)--(5.5,4)--(5.5,6)--(5.6,6)--(5.6,0)--(7.6,0)--(7.6,3.7)--(9,3.7)--(12,3.7)--(12,6.2)--(12.1,6.2)--(12.1,0.2)--(12.2,0.2)--(12.2,3.7)--(16.4,3.7)--(16.4,4.5)--(18,4.5)--(21.5,4.5)--(21.5,6)--(21.6,6)--(21.6,0)--(21.7,0)--(21.7,4.5)--(24,4.5);

		\draw[solid] (0,2.5)--(1,2.5)--(1,6)--(1.1,6)--(1.1,0)--(1.2,0)--(1.2,2.5)--(7,2.5)--(7,6.2)--(9.5,6.2)--(9.6,6.2)--(9.6,0.2)--(9.7,0.2)--(9.7,1.7)--(14.5,1.7)--(14.5,6.2)--(14.6,6.2)--(14.6,0.2)--(15.6,0.2)--(15.6,3.5)--(18,3.5)--(20,3.5)--(20,6)--(20.1,6)--(20.1,0)--(20.2,0)--(20.2,3.5)--(24,3.5)
		;
		
		\draw[solid] (0,1.5)--(1.5,1.5)--(1.5,6)--(1.6,6)--(1.6,0)--(1.7,0)--(1.7,1.5)--(6.6,1.5)--(6.6,4.7)--(9,4.7)--(12.5,4.7)--(12.5,6.2)--(12.6,6.2)--(12.6,0.2)--(12.7,0.2)--(12.7,4.7)--(16.8,4.7)--(16.8,6)--(18.5,6)--(18.6,6)--(18.6,0)--(18.7,0)--(18.7,2.5)--(23.5,2.5)--(23.5,6)--(23.6,6)--(23.6,0)--(24,0);
		
		
		\draw[solid] (0,5.5)--(6,5.5)--(6,5.4)--(2,5.4)--(2,0.5)--(0,0.5)--(0,0.4)--(6.2,0.4)--(6.2,2.7)--(9,2.7)--(10,2.7)--(10,6.2)--(10.1,6.2)--(10.1,0.2)--(10.2,0.2)--(10.2,2.7)--(16,2.7)--(16,1.5)--(18,1.5)--(22,1.5)--(22,6)--(22.1,6)--(22.1,0)--(22.2,0)--(22.2,1.5)--(24,1.5);
		
		\draw[solid] (0,4.5)--(2.5,4.5)--(2.5,6)--(2.6,6)--(2.6,0)--(2.7,0)--(2.7,4.5)--(7.8,4.5)--(7.8,5.7)--(9,5.7)--(15,5.7)--(15,5.6)--(13,5.6)--(13,0.7)--(9,0.7)--(9,0.6)--(15.2,0.6)--(15.2,1)--(18,1)--(20.5,1)--(20.5,6)--(20.6,6)--(20.6,0)--(20.7,0)--(20.7,1)--(24,1);
		
		\draw[solid] (0,3)--(3,3)--(3,6)--(3.1,6)--(3.1,0)--(3.2,0)--(3.2,3)--(7.2,3)--(7.2,4.2)--(9,4.2)--(10.5,4.2)--(10.5,6.2)--(10.6,6.2)--(10.6,0.2)--(10.7,0.2)--(10.7,4.2)--(16.6,4.2)--(16.6,5.5)--(18,5.5)--(24,5.5)--(24,5.4)--(19,5.4)--(19,0.5)--(18,0.5)--(18,0.4)--(24,0.4);
		
		\draw[solid] (0,2)--(3.5,2)--(3.5,6)--(3.6,6)--(3.6,0)--(3.7,0)--(3.7,2)--(6.8,2)--(6.8,2.2)--(9,2.2)--(13.5,2.2)--(13.5,6.2)--(13.6,6.2)--(13.6,0.2)--(13.7,0.2)--(13.7,2.2)--(15.8,2.2)--(15.8,5)--(18,5)--(22.5,5)--(22.5,6)--(22.6,6)--(22.6,0)--(22.7,0)--(22.7,5)--(24,5);
		
		\draw[solid] (0,1)--(4,1)--(4,6)--(4.1,6)--(4.1,0)--(4.2,0)--(4.2,1)--(6.4,1)--(6.4,5.2)--(9,5.2)--(11,5.2)--(11,6.2)--(11.1,6.2)--(11.1,0.2)--(11.2,0.2)--(11.2,5.2)--(17,5.2)--(17,4)--(18,4)--(21,4)--(21,6)--(21.1,6)--(21.1,0)--(21.2,0)--(21.2,4)--(24,4);
		
		\draw[solid] (0,5)--(4.5,5)--(4.5,6)--(4.6,6)--(4.6,0)--(4.7,0)--(4.7,5)--(8,5)--(8,3.2)--(9,3.2)--(14,3.2)--(14,6.2)--(14.1,6.2)--(14.1,0.2)--(14.2,0.2)--(14.2,3.2)--(16.2,3.2)--(16.2,3)--(18,3)--(19.5,3)--(19.5,6)--(19.6,6)--(19.6,0)--(19.7,0)--(19.7,3)--(24,3);
		
		\draw[solid] (0,3.5)--(5,3.5)--(5,6)--(5.1,6)--(5.1,0)--(5.2,0)--(5.2,3.5)--(7.4,3.5)--(7.4,1.2)--(9,1.2)--(11.5,1.2)--(11.5,6.2)--(11.6,6.2)--(11.6,0.2)--(11.7,0.2)--(11.7,1.2)--(15.4,1.2)--(15.4,2)--(18,2)--(23,2)--(23,6)--(23.1,6)--(23.1,0)--(23.2,0)--(23.2,2)--(24,2);
		%
		%
		%
		%
		%
		%
		%
		%
		
		
		\draw[solid,line width=1pt] (9.9,3.4)--(9.9,3)--(10.8,3);
		
		
		\draw[solid,line width=1pt] (5,2.2)--(5,1.4)--(5.3,1.4);
		
		
		\draw[solid,line width=1pt] (11,4.4)--(11,3.6)--(11.3,3.6);
		
		
		\draw[solid,line width=1pt] (20.4,3.2)--(20.4,2.8)--(21.3,2.8);
		
		\node at (-0.1,5.5) {4};
		\node at (-0.1,5) {9};
		\node at (-0.1,4.5) {5};
		\node at (-0.1,4) {1};
		\node at (-0.1,3.5) {10};
		\node at (-0.1,3) {6};
		\node at (-0.1,2.5) {2};
		\node at (-0.1,2) {7};
		\node at (-0.1,1.5) {3};
		\node at (-0.1,1) {8};
		\node at (-0.1,0.5) {4};
		
		\node at (6,5.5) {4};
		\node at (6,5) {9};
		\node at (6,4.5) {5};
		\node at (6,4) {1};
		\node at (6,3.5) {10};
		\node at (6,3) {6};
		\node at (6,2.5) {2};
		\node at (6,2) {7};
		\node at (6,1.5) {3};
		\node at (6,1) {8};
		\node at (6,0.5) {4};
		
		\node at (8.8,5.7) {5};
		\node at (8.8,5.2) {8};
		\node at (8.8,4.7) {3};
		\node at (8.8,4.2) {6};
		\node at (8.8,3.7) {1};
		\node at (8.8,3.2) {9};
		\node at (8.8,2.7) {4};
		\node at (8.8,2.2) {7};
		\node at (8.8,1.7) {2};
		\node at (8.8,1.2) {10};
		\node at (8.8,0.7) {5};
		
		\node at (15,5.7) {5};
		\node at (15,5.2) {8};
		\node at (15,4.7) {3};
		\node at (15,4.2) {6};
		\node at (15,3.7) {1};
		\node at (15,3.2) {9};
		\node at (15,2.7) {4};
		\node at (15,2.2) {7};
		\node at (15,1.7) {2};
		\node at (15,1.2) {10};
		\node at (15,0.7) {5};
		
		\node at (17.8,5.5) {6};
		\node at (17.8,5) {7};
		\node at (17.8,4.5) {1};
		\node at (17.8,4) {8};
		\node at (17.8,3.5) {2};
		\node at (17.8,3) {9};
		\node at (17.8,2.5) {3};
		\node at (17.8,2) {10};
		\node at (17.8,1.5) {4};
		\node at (17.8,1) {5};
		\node at (17.8,0.5) {6};
		
		\node at (24,5.5) {6};
		\node at (24,5) {7};
		\node at (24,4.5) {1};
		\node at (24,4) {8};
		\node at (24,3.5) {2};
		\node at (24,3) {9};
		\node at (24,2.5) {3};
		\node at (24,2) {10};
		\node at (24,1.5) {4};
		\node at (24,1) {5};
		\node at (24,0.5) {6};
		
		\node at (0.6,-0.2) {1};
		\node at (1.1,-0.2) {2};
		\node at (1.6,-0.2) {3};
		\node at (2.1,-0.2) {4};
		\node at (2.6,-0.2) {5};
		\node at (3.1,-0.2) {6};
		\node at (3.6,-0.2) {7};
		\node at (4.1,-0.2) {8};
		\node at (4.6,-0.2) {9};
		\node at (5.1,-0.2) {10};
		\node at (5.6,-0.2) {1};
		
		\node at (9.6,0) {2};
		\node at (10.1,0) {4};
		\node at (10.6,0) {6};
		\node at (11.1,0) {8};
		\node at (11.6,0) {10};
		\node at (12.1,0) {1};
		\node at (12.6,0) {3};
		\node at (13.1,0) {5};
		\node at (13.6,0) {7};
		\node at (14.1,0) {9};
		\node at (14.6,0) {2};
		
		\node at (18.6,-0.2) {3};
		\node at (19.1,-0.2) {6};
		\node at (19.6,-0.2) {9};
		\node at (20.1,-0.2) {2};
		\node at (20.6,-0.2) {5};
		\node at (21.1,-0.2) {8};
		\node at (21.6,-0.2) {1};
		\node at (22.1,-0.2) {4};
		\node at (22.6,-0.2) {7};
		\node at (23.1,-0.2) {10};
		\node at (23.6,-0.2) {3};

		\node at (0.5,6.2) {1};
		\node at (1,6.2) {2};
		\node at (1.5,6.2) {3};
		\node at (2,6.2) {4};
		\node at (2.5,6.2) {5};
		\node at (3,6.2) {6};
		\node at (3.5,6.2) {7};
		\node at (4,6.2) {8};
		\node at (4.5,6.2) {9};
		\node at (5,6.2) {10};
		\node at (5.5,6.2) {1};
		
		\node at (9.6,6.4) {2};
		\node at (10,6.4) {4};
		\node at (10.5,6.4) {6};
		\node at (11,6.4) {8};
		\node at (11.5,6.4) {10};
		\node at (12,6.4) {1};
		\node at (12.5,6.4) {3};
		\node at (13,6.4) {5};
		\node at (13.5,6.4) {7};
		\node at (14,6.4) {9};
		\node at (14.5,6.4) {2};
		
		\node at (18.5,6.2) {3};
		\node at (19,6.2) {6};
		\node at (19.5,6.2) {9};
		\node at (20,6.2) {2};
		\node at (20.5,6.2) {5};
		\node at (21,6.2) {8};
		\node at (21.5,6.2) {1};
		\node at (22,6.2) {4};
		\node at (22.5,6.2) {7};
		\node at (23,6.2) {10};
		\node at (23.5,6.2) {3};
		
		\end{tikzpicture}
		\caption{Part of $PB_{2n+4}$-VPG representation of $K_{10}^3$.}
		\label{fig:properbend2n+4}
	\end{sidewaysfigure}

	\paragraph{Example:}
	\sloppy Here we illustrate the process stated in the proof of Lemma~\ref{lem:ubound-2} by constructing a proper $B_{24}$-VPG representation of $K^3_{10}$. Notice that $P(a_2)$ has $24$ bends. By adding three dummy vertices to vertex set $\{a_i:i\in[10]\}$, we construct a complete graph on $13$ vertices, and find six edge disjoint Hamiltonian cycles. The six sequences obtained from them, after removing the dummy vertices, are as follows: $S_1=(1,2,3,4,5,6,7,8,9,10,1)$,  $S_2=(2,4,6,8,10,1,3,5,7,9,2)$, $S_3=(3,6,9,2,5,8,1,4,7,10,3)$,  $S_4=(4,8,3,7,2,6,10,1,5,9,4)$, $S_5=(5,10,2,7,4,9,1,6,3,8,5)$ and $S_6=(6,5,4,10,3,9,2,8,1,7,6)$. So we will have three square regions; the first representing $S_1$ and $S_4$, the second representing $S_2$ and $S_5$, and the third representing $S_3$ and $S_6$. Using the three operations defined in the procedure, we complete the representations in each square region. In the gaps between two consecutive square regions, each path has to change its $Y$-coordinate, which it can do using two bends. To ensure a proper intersection in these gaps, the $X$-coordinate of these bends in each path are distinct. See Fig. \ref{fig:properbend2n+4} for a partial representation. We show $P(b)$'s (in bold) for the $b$'s that are adjacent to $(a_3,a_7,a_{10})$, $(a_4,a_6,a_{9})$, $(a_1,a_6,a_{8})$ and $(a_5,a_8,a_{9})$.

	\subsection{Proof of Lemma~\ref{lem:bendub}}\label{sec:lem:bendub}
	Now we shall give a $B_{4t+29}$-VPG representation of $G_{t,m}$, in fact we give its $PB_{4t+29}$-VPG representation. First we introduce the following definitions. Recall that $k \geq 2t+16$ and $n= 2k^2k!+3$.
	
	Let $\mathcal{R}$ be a $B_l$-VPG representation of a $B_l$-VPG graph $G$ and $P(v)$ be a path in $\mathcal{R}$ corresponding to a vertex $v\in V(G)$. \emph{Corner points} of a $k$-bend path are the end points of the horizontal and vertical segments of the path. Let $p_1,\ldots,p_r$ be the corner points encountered while traversing the the path $P(v)$ from $p_1$ to $p_r$, where $p_1,p_r$ are the endpoints of $P(v)$. The \emph{direction vector} of $P(v)$ is a vector of size $r-1$ where each entry is a symbol from $\{\rightarrow,\leftarrow,\downarrow,\uparrow\}$. For some $i\in \{1,2,\ldots,r-1\}$, let $f_i$ denote the $i^{th}$ entry of the direction vector of $P(v)$ in $\mathcal{R}$. If $f_i$ is $\rightarrow$, then it means that $p_{i+1}$ lies horizontally right of $p_i$; if $f_i$ is $\leftarrow$, then it means that $p_{i+1}$ lies horizontally left of $p_i$; if $f_i$ is $\downarrow$, then it means that $p_{i+1}$ lies vertically below $p_i$; and if $f_i$ is $\uparrow$, then it means that $p_{i+1}$ lies vertically above $p_i$.
	
\begin{figure}
		\centering
		\begin{tabular}{cccc}
			&     
			\begin{tikzpicture}[scale=1.2]
			
			\fill [fill=black!40!white,opacity=0.3] (-0.5,0.25) rectangle (2,0.5);
			\draw [solid] (0,1.5) -- (0,0.75);
			\draw [thick, dotted] (0,0.5) -- (0,0.75);
			\draw [thick, dotted] (0,0.5) -- (0.25,0.5);
			\draw [solid] (1,0.5) -- (0.25,0.5);
			
			\draw [solid] (0.25,1.25) -- (0.25,0.5);
			\draw [thick, dotted] (0.25,0.25) -- (0.25,0.5);
			\draw [thick, dotted] (0.25,0.25) -- (0.5,0.25);
			\draw [solid] (1.25,0.25) -- (0.5,0.25);
			
			\draw [solid] (0.5,1) -- (0.5,0.25);
			\draw [thick, dotted] (0.5,0) -- (0.5,0.25);
			\draw [thick, dotted] (0.5,0) -- (0.75,0);
			\draw [solid] (1.5,0) -- (0.75,0);
			
			\draw[->,solid] (-0.3,0.4)--(-0.3,-0.1);
			\node at (0.2,-0.3) {\scriptsize Exposed Zone};
			\end{tikzpicture}
			& \\
			&(a)&\\
			\vspace{5pt}
			\begin{tikzpicture}[scale=0.73]
			\draw [dashed] (2,2) -- (2.5,2);
			\draw [solid] (2.5,2) -- (5.5,2);
			\draw [dashed] (5.5,2) -- (6,2);
			\draw [solid] (4,2.6)--(4,0.5);
			\draw [<->,solid] (4.2,2.6)--(4.2,2);
			\draw[fill] (4,2.6) circle (1pt);
			\draw[->,dotted] (4,2.6)--(3.5,3.3);
			
			\node at (4.4,2.3) {$\epsilon_t$};
			\node at (6,2.28) {$P(a_i)$};
			\node at (3.4,2.5) {$P(b_j)$};
			\node at (3.5,3.5) {\scriptsize Starting point of $P(b_j)$};
			\end{tikzpicture}
			& \begin{tikzpicture}[scale=0.73]
			\fill [fill=black!40!white,opacity=0.3] (1,0) rectangle (6,2);
			\draw [<->,solid] (6,0)--(6,2);
			\draw [<->,solid] (3.2,2)--(3.2,1.5);
			\draw [solid] (3.5,2.5)--(3.5,1.5)--(4,1.5);
			\draw[dashed] (4,1.5)--(5,1.5);
			\draw[dashed] (3.5,2.5)--(3.5,3.5);
			\draw[->,dotted] (2,1)--(2,2.8);
			
			\node at (6.3,1) {$\epsilon_0$};
			\node at (3,1.5) {$\epsilon_t$};
			\node at (3.6,2.5) {$P(b_j)$};
			\node at (2,3) {\scriptsize Exposed Zone};
			\end{tikzpicture}
			& \begin{tikzpicture}[scale=0.73]
			\draw [dashed] (2,2) -- (3,2);
			\draw [solid] (3,2) -- (6,2) -- (6,1.5);
			\draw [dashed] (6,1.5) -- (6,0.5);
			\draw [dashed] (5,4) -- (5,3.5);
			\draw [solid] (5,3.5)--(5,1);
			\draw [dashed] (5,0.5)--(5,1);
			\draw[<->,solid] (5,2.2) -- (6,2.2);
			\node at (5.5,2.5) {$\epsilon_t$};
			\node at (5.8,3.5) {$P(a_i)$};
			\node at (2.5,2.5) {$P(b_j)$};
			\end{tikzpicture}
			\\
			(b) & (c) & (d)
		\end{tabular}
		\caption{(a) Exposed zone is highlighted in grey. (b) shows the starting point of $P(b_j)$. (c) shows the corner point when $P(b_j)$ just enters an exposed zone and (c) shows the corner point when $P(b_j)$ crosses some $P(a_i)$.}
		\label{fig:just}
	\end{figure}
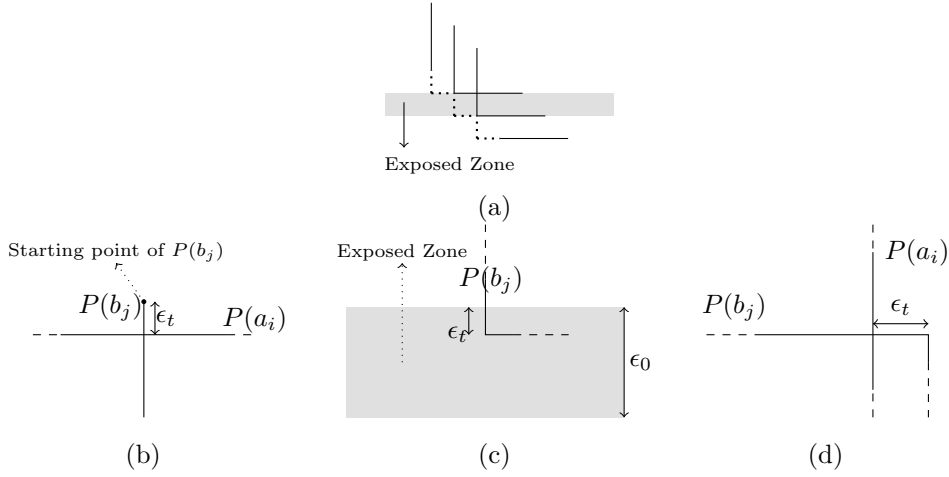

	We call a part of a horizontal (vertical) segment of $P(v)$ is \emph{exposed from below} (resp. \emph{exposed from left}) in $\mathcal{R}$, if a vertical (resp. horizontal) line drawn downwards (resp. leftwards) from any point in this part does not intersect with any path $P(u)$ in $\mathcal{R}$ where $u\in V(G)$. \emph{Exposed parts} of the path $P(v)$ are the parts of segments which are either exposed from below or exposed from left. For vertical segments of $P(v)$ we define its \emph{exposed zone} as the infinite horizontal strip just containing the exposed part of this segment (see Figure~\ref{fig:just}(a)). Below we give a $PB_{2k-3}$-VPG representation of $G_{t,m}$. (Notice that $2k-3=4t+29$.)
	
	\sloppy Recall that the vertex set of $G_{t,m}$ is $A\cup Q$ where $A = \{ a_i \mid i \in [n]  \}$ and $Q=\{b_j \mid j\in \binom{n}{k}\}$. For $i\in [n]$, $a_i$ is indexed by $[n]\backslash \{i\}$; and for $j\in \binom{n}{k}$, $b_j$ is indexed by a distinct $k$ element subset of $[n]$. The edge set $E(G_{t,m})$ $= \{a_ia_j \mid i\neq j,\forall a_i, a_j \in A\} \cup \{ b_ib_j \mid i\neq j,\forall b_i, b_j \in Q \} \cup \{a_ib_j \mid i \text{ belongs to the $k$-element subset of $[n]$ indexed by $b_j$}\}$.
	
	First we describe how to represent the vertices of $A=\{a_1,a_2,\ldots,a_n\}$. We represent $a_1$ by a 3 bend stair $P(a_1)$ with direction vector $<\rightarrow, \downarrow, \rightarrow, \downarrow  >$ with segments of unit length. Let its starting point (first corner point) be the origin, with the usual positive X $(\rightarrow)$ and positive Y $(\uparrow)$ axis orientation. Fix $\epsilon_0 << 1$. Now we draw $P(a_i)$, for each $i\in \{2,3,\ldots n\}$. Each such $P(a_i)$ is a congruent copy of $P(a_1)$ starting from $\left((i-1)\epsilon_0,-(i-1)\epsilon_0\right)$ with the same direction vector as $P(a_1)$. The $a_i$'s, for $i \in [n]$, induce a clique; as for distinct $a_i$ and $a_j$ with $i<j$, the first bend i.e. second corner point of $P(a_i)$ occurs to the left of the second corner point of $P(a_j)$ and the first segment of $P(a_i)$ lies above the first segment of $P(a_j)$, so the second segment of $P(a_i)$ intersects first segment of $P(a_j)$. Let $\mathcal{R}_A$ denote the above VPG representation of vertices in $A$.

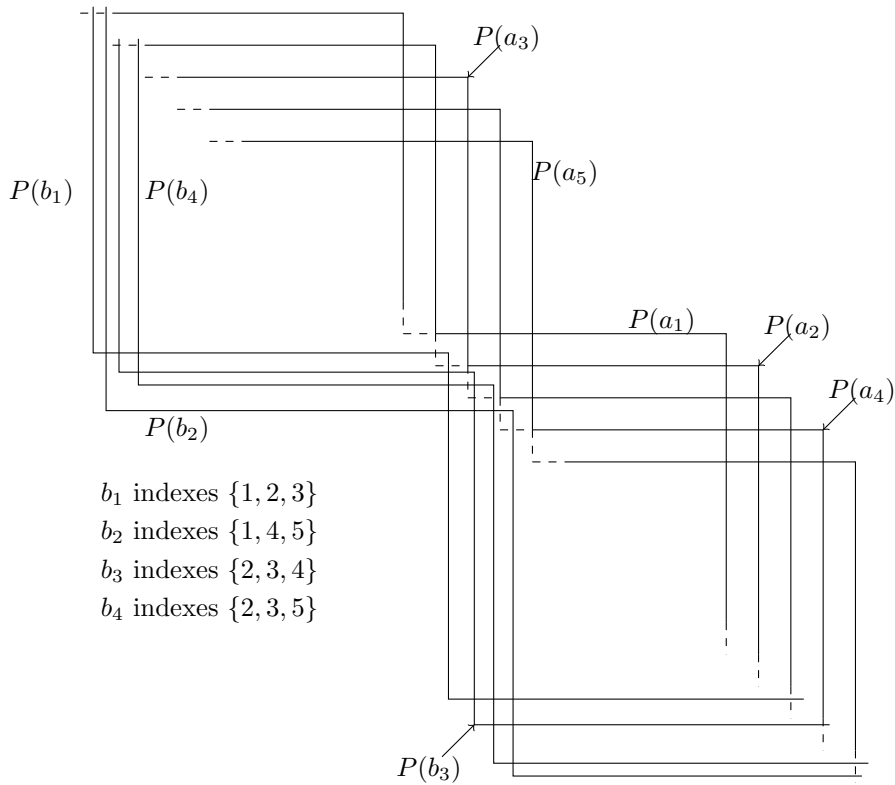
\begin{figure}
		\begin{tikzpicture}[scale=0.85]
		
		\draw [dashed] (0,10)--(0.5,10);
		\draw [solid] (0.5,10)--(5,10)--(5,5.5);
		\draw [dashed] (5,5.5)--(5,5);
		\draw [dashed] (5,5)--(5.5,5);
		\draw [solid] (5.5,5)--(10,5)--(10,0.5);
		\draw [dashed] (10,0.5)--(10,0);
		
		\draw [dashed] (0.5,9.5)--(1,9.5);
		\draw [solid] (1,9.5)--(5.5,9.5)--(5.5,5);
		\draw [dashed] (5.5,5)--(5.5,4.5);
		\draw [dashed] (5.5,4.5)--(6,4.5);
		\draw [solid] (6,4.5)--(10.5,4.5)--(10.5,0);
		\draw [dashed] (10.5,0)--(10.5,-0.5);
		
		\draw [dashed] (1,9)--(1.5,9);
		\draw [solid] (1.5,9)--(6,9)--(6,4.5);
		\draw [dashed] (6,4.5)--(6,4);
		\draw [dashed] (6,4)--(6.5,4);
		\draw [solid] (6.5,4)--(11,4)--(11,-0.5);
		\draw [dashed] (11,-0.5)--(11,-1);    
		
		\draw [dashed] (1.5,8.5)--(2,8.5);
		\draw [solid] (2,8.5)--(6.5,8.5)--(6.5,4);
		\draw [dashed] (6.5,4)--(6.5,3.5);
		\draw [dashed] (6.5,3.5)--(7,3.5);
		\draw [solid] (7,3.5)--(11.5,3.5)--(11.5,-1);
		\draw [dashed] (11.5,-1)--(11.5,-1.5);
		
		\draw [dashed] (2,8)--(2.5,8);
		\draw [solid] (2.5,8)--(7,8)--(7,3.5);
		\draw [dashed] (7,3.5)--(7,3);
		\draw [dashed] (7,3)--(7.5,3);
		\draw [solid] (7.5,3)--(12,3)--(12,-1.5);
		\draw [dashed] (12,-1.5)--(12,-2);

		
		\draw[solid] (0.2,10.1)--(0.2,4.7)--(5.7,4.7)--(5.7,-0.7)--(11.2,-0.7);
		\draw[solid] (0.4,10.1)--(0.4,3.8) --(6.7,3.8)--(6.7,-1.9)--(12.1,-1.9);
		\draw[solid] (0.6,9.6)--(0.6,4.4)--(6.1,4.4)--(6.1,-1.1)--(11.6,-1.1);
		\draw[solid] (0.9,9.6)--(0.9,4.2)--(6.4,4.2)--(6.4,-1.7)--(12.2,-1.7);
		
		\draw [<-] (6,9)--(6.5,9.5);
		\draw [<-] (10.5,4.5)--(11,5);
		\draw [<-] (11.5,3.5)--(12,4);
		\draw [<-] (6.1,-1.1)--(5.6,-1.6);
		
		\node at (9,5.2) {$P(a_1)$};
		\node at (11.1,5.1) {$P(a_2)$};
		\node at (6.6,9.6) {$P(a_3)$};
		\node at (12.1,4.1) {$P(a_4)$};
		\node at (7.5,7.5) {$P(a_5)$};
		\node at (-0.6,7.2) {$P(b_1)$};
		\node at (1.5,7.2) {$P(b_4)$};
		\node at (1.5,3.5) {$P(b_2)$};
		\node at (5.4,-1.8) {$P(b_3)$};
		\node at (2,2.5) {$b_1 ~\text{indexes } \{1,2,3\}$};
		\node at (2,1.9) {$b_2 ~\text{indexes } \{1,4,5\}$};
		\node at (2,1.3) {$b_3 ~\text{indexes } \{2,3,4\}$};
		\node at (2,0.7) {$b_4  ~\text{indexes}~ \{2,3,5\}$};
		\end{tikzpicture}
		\caption{Part of $B_{2k-3}$-VPG representation with $n=5,k=3$ (for illustration).}\label{fig:example}
	\end{figure}

	Observe that, for each $i\in [n]$, the first $\epsilon_0$ length of the first and third segment of $P(a_i)$ i.e. $\epsilon_0$ length of the first and third segment starting from the first and third corner point respectively, are exposed in $\mathcal{R}_A$. Similarly the end $\epsilon_0$ length of second and fourth segment of $P(a_i)$ i.e. $\epsilon_0$ length of the second and fourth segment ending at the third and fifth corner point respectively, are also exposed in $\mathcal{R}_A$ (see Figure~\ref{fig:just}(a) and \ref{fig:example}).

	Now for each $b_i \in Q$, for $i \in \binom{n}{k}$, we add $2k-3$ bend stair $P(b_i)$ with direction vector $<\downarrow, \rightarrow, \downarrow, \rightarrow, \ldots, \downarrow, \rightarrow>$. Assume $\{i_1, i_2,\ldots, i_k\}$ is indexed by $b_i$ with $i_1<i_2<\ldots<i_k$ such that the vertex $b_i$ is adjacent to the vertices $\{a_{i_1},a_{i_2},\ldots,a_{i_k}\}$. When we say $P(b_i)$ starts \emph{just above} a segment, or \emph{just enters} an exposed zone, or \emph{just crosses} a path, we mean that $P(b_i)$ starts $\epsilon_t$ distance above a segment, enters $\epsilon_t$ distance into an exposed zone, or crosses and goes on for $\epsilon_t$ distance respectively for some $\epsilon_t < \epsilon_0$ (see Figure~\ref{fig:just}(b),(c),(d)). Also $\epsilon_t$ is assumed to take distinct values so as to result in a  $PB_{2k-3}$-VPG representation. Next we describe how to draw $P(b_i)$.
	
	The first segment of $P(b_i)$ starts just above the exposed part of the first segment of $a_{i_1}$. For two vertices $b_j, b_l \in B$, where $b_j$ indexes $\{j_1, j_2,\ldots, j_k\}$ and $b_l$ indexes $\{l_1, l_2,\ldots, l_k\}$, if $j_1 = l_1$, the horizontal positions of their starting points are lexicographically ordered from left to right. The first bend occurs when $P(b_i)$ just enters the exposed zone of the second segment of $P(a_{i_2})$. For two vertices $b_j, b_l \in B$ with $j_2 = l_2$, the first bends i.e. second corner points are lexicographically ordered from top to bottom. The second bend occurs after $P(b_i)$ just crosses $P(a_{i_2})$. Till now $P(b_i)$ has intersected with $P(a_{i_1})$ and $P(a_{i_2})$. For rest of the bends in $P(b_i)$ we give a general scheme. For intersection with $P_{a_r}$ for $r\in \{3,4,\ldots,k-1\}$, the $2r-3^{\text{th}}$ bend occurs when $P(b_i)$ just enters the exposed zone of the fourth segment of $P(a_{i_r})$ and the $2r-2^{\text{th}}$ bend occurs after $P(b_i)$ just crosses $P(a_{i_r})$. For intersection with $P_{a_k}$, the $2k-3^{\text{th}}$ bend occurs when $P(b_i)$ just enters the exposed zone of the fourth segment of $P(a_{i_k})$ and ends when $P(b_i)$ just crosses $P(a_{i_k})$ (see Figure~\ref{fig:example}).
	
	Notice that $P(b_i)$, for $i \in [\binom{n}{k}]$, crosses $P(a_{i_1})$ and the bends in $P(b_i)$ occur when it either enters an exposed zone of $P(a_{i_r})$, for $r = 2$ to $k$, or just after crossing $P(a_{i_r})$ for $r = 2$ to $k-1$. It ends just after crossing $P(a_{i_k})$. Hence $P(b_i)$ intersects only $P(a_{i_1})$, $P(a_{i_2})$, \ldots, $P(a_{i_k})$. Also every $P(b_i)$ has $2k-3$ bends. Now we prove that $b_i$'s, for $i \in [\binom{n}{k}]$, induce a clique. Consider two vertices $b_i$ and $b_j$ with $i \neq j$ where $b_i$ indexes $\{i_1, i_2,\ldots, i_k\}$ and $b_j$ indexes $\{j_1, j_2,\ldots, j_k\}$ such that $i_1<i_2<\ldots<i_k$ and $j_1<j_2<\ldots<j_k$. Without loss of generality assume $\{i_1, i_2,\ldots, i_k\}$ is lexicographically less than $\{j_1, j_2,\ldots, j_k\}$. Clearly $i_1 \leq j_1$. We prove that $P(b_i)$ and $P(b_j)$ intersect. Recall that, due to the lexicographic ordering constructed in the previous paragraph, if $i_1 = j_1$, then the starting point of $P(b_i)$ lies to the left of that of $P(b_j)$; and if $i_2=j_2$, then the first bend of $P(b_i)$ lies above that of $P(b_j)$.
	
	\begin{enumerate}
		\renewcommand{\theenumi}{(\roman{enumi})}
		\renewcommand{\labelenumi}{\theenumi}
		
		\item If $i_2 < j_2$, the first bend of $P(b_i)$ occurs above the first bend of $P(b_j)$ and the starting point of $P(b_i)$ is to the left of the starting point of $P(b_j)$ (even if $i_1=j_1$, due to the lexicographic ordering); hence the second segment of $P(b_i)$ intersects the first segment of $P(b_j)$. (See paths $P(b_1)$ and $P(b_2)$ in Figure~\ref{fig:example}.)
		
		\item If $i_2 = j_2$, the first bend point of $P(b_i)$ occurs above the first bend of $P(b_j)$ (due to the lexicographic ordering) and the starting point of $P(b_i)$ is to the left of the starting point of $P(b_j)$; hence second segment of $P(b_i)$ intersects the first segment of $P(b_j)$. (See paths $P(b_3)$ and $P(b_4)$ in Figure~\ref{fig:example}.)
		
		\item If $i_2 > j_2$, then $i_1 < j_1$ (else it contradicts the lexicographic ordering between $\{i_1, i_2,\ldots, i_k\}$ and $\{j_1, j_2,\ldots, j_k\}$). In such a case the first bend point of $P(b_i)$ occurs below that of $P(b_j)$ and the second bend point of $P(b_i)$ occurs to the right of that of $P(b_j)$; hence the second segment of $P(b_i)$ intersects the third segment of $P(b_j)$. (See paths $P(b_2)$ and $P(b_3)$ in Figure~\ref{fig:example}.)
		
	\end{enumerate}
	
	So $P(b_i)$ and $P(b_j)$, for $i, j \in [\binom{n}{k}]$, intersect and hence $b_i$'s, for $i \in [\binom{n}{k}]$, induce a clique. This completes the $PB_{2k-3}$-VPG representation i.e. $B_{4t+29}$-VPG representation of $G_{t,m}$.

	\section{Conclusions}\label{sec:conclude}
	
	This article primarily focuses on constructing cocomparability graphs with low bend number and high poset dimension. One further direction would be to study the bend number of asteroidal-triple free graphs which is a superclass of cocomparability graphs.

	We also aimed to separate $B_k$-VPG-chordal and $B_{k+1}$-VPG-chordal for every $k\in \mathbb{N}$. To this end we studied bend numbers of subclasses of split graphs, $Forb(C_{\geq 5})$ graphs and cocomparability graphs. We proved that there are infinitely many $m\in \mathbb{N}$ such that $B_{m}$-VPG-split $\subsetneq$ $B_{m+1}$-VPG-split. We also prove for every $t\in \mathbb{N}$, $B_{t}$-VPG-$Forb(C_{\geq 5})$ $\subsetneq$ $B_{4t+29}$-VPG-$Forb(C_{\geq 5})$.  The original question of whether $B_k$-VPG-chordal $\subsetneq B_{k+1}$-VPG-chordal for every $k\in \mathbb{N}$ however remains open. The following question might be interesting in this direction.
	
	\begin{question}
		Let $\mathcal{C}$ denote any graph class among $\textsc{split, chordal, $Forb(C_{\geq 5})$}$. Are there separating examples such that $B_k$-VPG-$\mathcal{C} \subsetneq B_{k+1}$-VPG-$\mathcal{C}$ and $PB_k$-VPG-$\mathcal{C} \subsetneq PB_{k+1}$-VPG-$\mathcal{C}$ ?
	\end{question}
	
	We believe that answering the above question would lead to interesting techniques to prove lower bounds on bend number and proper bend number of string graphs. We proved that $PB_{t}$-VPG-split $\subsetneq$ $PB_{36t+80}$-VPG-split. Since $bend(G)$ is always less than $bend_p(G)$ the following question is interesting.
	
	\begin{question}
		For any string graph $G$, does $bend_p(G)$ lie within a constant additive (multiplicative) factor of $bend(G)$?
	\end{question}
	
	When a graph $G$ is a split graph, we gave an upper bound on $bend(G)$ (see Theorem~\ref{thm:uperbound-split}). To the best of our knowledge, this is the first non-trivial upper bound on the bend number of split graphs.
	
	\section*{Acknowledgements}
	
	We thank Sagnik Sen for helpful comments in preparing the manuscript. We thank Subhodeep Ranjan Dev for carefully reading the final draft. We would like to thank the anonymous referees for meticulously reading the manuscript, and for helpful suggestions which made the proofs more rigorous and increased the readability and flow of the manuscript. Joydeep Mukherjee is supported by DST SERB NPDF fellowship (PDF/2016/001647).


	

	\bibliographystyle{plain}      
	\bibliography{reference}

\end{document}